\documentclass[twocolumn]{aastex631}

\newcommand{\kms}{km~s$^{-1}$\,}

\newcommand{\msun}{${\cal M}_\odot$\,}

\begin{document}

\renewcommand{\topfraction}{1.0}
\renewcommand{\bottomfraction}{1.0}
\renewcommand{\textfraction}{0.0}

\shorttitle{Speckle Interferometry at SOAR}

\title{Speckle Interferometry at SOAR in 2023}

\author{Andrei Tokovinin}
\affil{Cerro Tololo Inter-American Observatory | NFSs NOIRLab Casilla 603, La 
Serena, Chile}
\email{andrei.tokovinin@noirlab.edu}

\author{Brian D. Mason}
\affil{U.S.\ Naval Observatory, 3450 Massachusetts Ave., Washington, DC, USA}
\email{brian.d.mason.civ@us.navy.mil}

\author{Rene A. Mendez}
\affil{Universidad de Chile, Casilla 36-D, Santiago, Chile}
\email{rmendez@uchile.cl}
\author{Edgardo Costa}
\affil{Universidad de Chile, Casilla 36-D, Santiago, Chile}

\begin{abstract}
Results  of  the  speckle-interferometry  observations at  the  4.1  m
Southern Astrophysical Research Telescope  (SOAR) obtained during 2023
are presented: 1913 measurements of  1533 resolved pairs or subsystems
(median separation  0\farcs16) and non-resolutions of  552 targets; 42
pairs are  resolved here for the  first time. This work  continues our
long-term  effort to  monitor  orbital motion  in  close binaries  and
hierarchical  systems.   A large  number  (147)  of orbits  have  been
determined   for    the   first   time   or    updated   using   these
measurements. Complementarity of this program with the Gaia mission is
highlighted.
\end{abstract} 

\section{Introduction}
\label{sec:intro}

This paper continues the series of double-star measurements made at the 4.1 m 
SOuthern Astrophysical Research Telescope (SOAR) since 2008 with the speckle 
camera, HRCam. Previous results are published by \citet[][hereafter 
TMH10]{TMH10} and in \citep{SAM09,Hrt2012a,Tok2012a,TMH14,TMH15,SAM15,SAM17,
SAM18,SAM19,SAM20,SAM21,SAM22}. Observations reported here were made
during 2023.

\begin{figure}[ht]
\epsscale{1.1}
\plotone{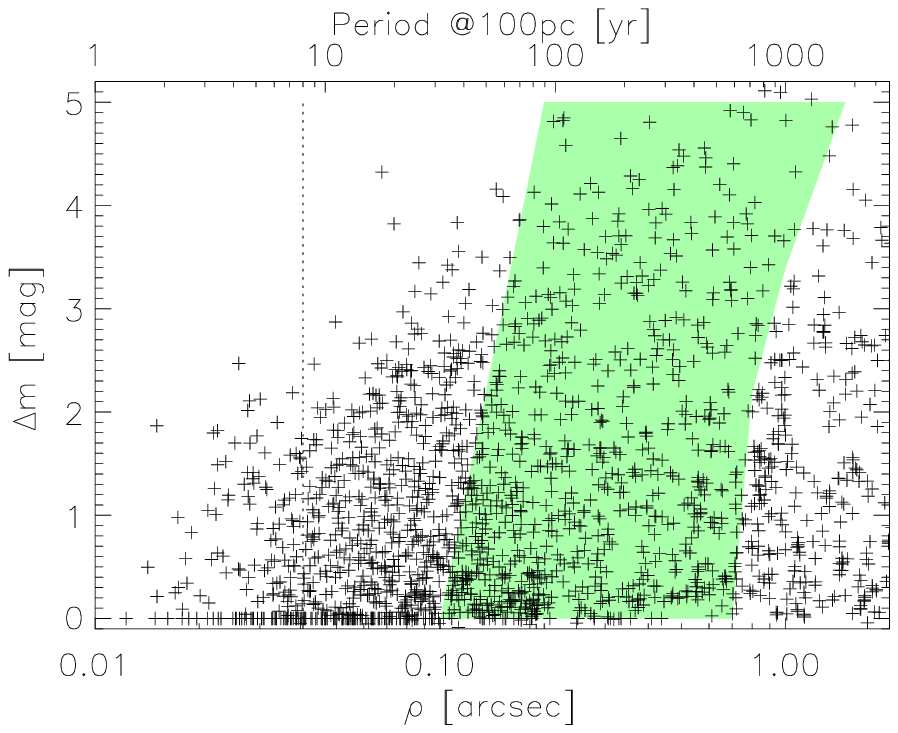}
\caption{Magnitude difference  $\Delta m$  vs.  separation  $\rho$ for
  pairs measured in  2023 (crosses). The upper  axis indicates periods
  of binaries  with such  separations and  a mass sum  of 1  \msun ~if
  located at  100\,pc distance.  The hatched  area highlights expected
  future measurements  from Gaia, the  vertical dotted line  shows the
  SOAR diffraction limit in the $I$ band.
\label{fig:dmsep} }
\end{figure}

The Gaia  space mission \citep{Gaia1}  is having a profound  impact in
many areas, including binary stars, so  it is appropriate to place our
ongoing  program   in  this  context.    Figure~\ref{fig:dmsep}  plots
separations and  magnitude differences  of pairs  measured at  SOAR in
2023.  The  green shading indicates  pairs expected to be  resolved by
the 1\,m  Gaia apertures at  the diffraction limit of  0\farcs1; their
individual measurements will become available in future data releases,
while for wider pairs to the right of the shaded zone the positions of
both components are already available in the current Gaia data release
3  \citep[GDR3,][]{EDR3}.    At  a   distance  of   100\,pc,  0\farcs1
separation corresponds to orbital periods on the order of 30\,yr which
are common for low-mass binaries.  Ground-based speckle interferometry
is the only source  of data for tracing orbits of  such pairs, and our
program makes a significant  contribution here.  The expected duration
of  the  full  Gaia  mission   will  not  suffice  for  derivation  of
astrometric  or   spectroscopic  orbits   with  periods   longer  than
$\sim$10\,yr, but  the combination of  the Gaia and  ground-based data
opens  exciting  perspectives  for accurate  measurements  of  stellar
masses and for other applications.

The structure and content of this paper are similar to other papers of this 
project. Section~\ref{sec:obs} reviews all speckle programs that contributed to 
this paper, the observing procedure, and the data reduction. The results are 
presented in Section~\ref{sec:res} in the form of electronic tables archived by 
the journal. We also discuss new resolutions and present orbits resulting from
this data set. A short summary and an outlook of further work in 
Section~\ref{sec:sum} close the paper.

\section{Observations}
\label{sec:obs}

\subsection{Observing Programs}

As  in previous  years,  HRCam (see  Section~\ref{sec:inst}) was  used
during 2023  to execute several  observing programs, some  with common
targets.  In the  data tables, new or recently  discovered objects are
linked to respective  programs by labels in place  of usual discoverer
designations   (that    are   not    yet   assigned),    see   Section
\ref{sec:tables}. They are also marked by tags, where N means new pair
resolved in  2023, and other  tags refer to pairs  resolved previously
but not yet published.

{\it  Hierarchical systems}  are currently  the core  of our  program.
Their  architecture is  relevant  to star  formation, while  dynamical
evolution   of  these   hierarchies  increases   chances  of   stellar
interactions  and  mergers  \citep{review}. Orbital  motions  of  many
triple systems  are monitored  at SOAR,  and these  data are  used for
orbit        determinations        and       dynamical        analysis
\citep{TL2020,Tok2021a,CHI23,TRI23}.  We re-observed  some tight inner
subsystems  of   hierarchies  discovered  at  SOAR   in  2021-2022  by
\citet[][label POW, tag P]{Powell2023}  and hierarchies within 100\,pc
\citet[][label  GKM, tag  T]{GKM}, see  Section \ref{sec:GKM}.   Wider
pairs within 100 pc  resolved by Gaia but not yet  featured in the WDS
were  observed in  search  for  inner subsystems  (tag  G).  Also,  on
request  by   S.~Majewski,  we   targeted  several   doubly  eclipsing
quadruples discovered from the  TESS photometry by \citet{Kostov2022};
these stars  are labeled  QUAD in  the data tables.   New data  on all
hierarchical systems are  reflected in the Multiple  Star Catalog, MSC
\citep{MSC}.  Its latest  update in 2023 December is  available at the
MSC   \href{https://www.ctio.noirlab.edu/~atokovin/stars/}{home  page}
and                                                                 at
\href{http://vizier.u-strasbg.fr/viz-bin/VizieR-4?-source=J/ApJS/235/6}{Vizier}.

{\it Orbits} of resolved binaries are improved in quantity and quality
using     our     new     measurements,    contributing     to     the
\href{https://crf.usno.navy.mil/wds-orb6/}{Sixth Catalog  of Orbits of
  Visual  Binary  Stars}  \citep{VB6}.   We provide  large  tables  of
reliable and  preliminary orbits in Section  \ref{sec:orbits}. Data on
visual  orbits  serve in  many  areas,  e.g.   to probe  alignment  of
exoplanetary orbits \citep{Lester2023}.

{\it Hipparcos binaries} within 200\,pc are monitored to measure masses of stars
and to test stellar evolutionary models, as outlined by, e.g., 
\citet{Horch2015,Horch2017,Horch2019}. The southern part of this sample is 
addressed at SOAR \citep{Mendez2017}. This program overlaps with the general 
work on visual orbits.

{\it Nearby M dwarfs} are being  observed at SOAR since 2018 following
the initiative of T.\ Henry and  E.\ Vrijmoet. The goal is to assemble
statistical data on orbital elements, focusing on short periods. First
results on M dwarfs are published by \citet{Vrijmoet2022}. In 2023, we
continued to monitor  these pairs; a paper  containing $\sim$50 orbits
of M dwarfs  is in preparation. In anticipation, residuals  to some of
those orbits are given in  the data table.  Measurements of previously
known  pairs  are  published  here,   and  those  of  new  pairs  from
\citet{Vrijmoet2022} are deferred to the  paper on orbits.  In 2023, a
modest  number of  additional  low-mass  M dwarfs  within  25 pc  were
targeted (labeled as M25), and two were resolved.

{\it  Neglected  close  binaries}  from  the  Washington  Double  Star
Catalog,      WDS       \citep{WDS},\footnote{See      the      latest
  \href{https://crf.usno.navy.mil/wds/}{online}  WDS   version.}  were
observed as a ``filler'' at low  priority. 

{\it TESS follow-up} continues the  program executed in 2018--2020, at
a reduced rate.  Its past results  are published by \citet{TESS,TESS2}. All
speckle observations of  TESS targets of interest  are promptly posted
on the  \href{https://exofop.ipac.caltech.edu/tess/}{EXOFOP web site.}
These  data  are  used  in  the  growing  number  of  papers  on  TESS
exoplanets, mostly  as limits  on close  companions to  exohosts. We
publish here measures of four pairs from the TESS program resolved for
the first time in 2023 and lacking relative positions in  GDR3 (tag Z). 

{\it  Acceleration  stars} were  observed  in  2021-2022 as  potential
targets of  high-contrast imaging  of exoplanets in  a program  led by
K.\ Franson and B.\ Bowler (tag A). While their paper is still in preparation,
we  re-observed  many of  resolved  targets  in  2023 to  confirm  the
detections and  to clarify  their nature.  

A few close subsystems in {\em  wide pairs} resolved previously in the
program led  by J.\ Chanam\'e (tag C)  have been   re-observed in  2023 to
detect  orbital  motion, while  their  first  resolutions still  await
publication.

Speckle observations  in 2023 were  conducted during 9  observing runs
for a  total of approximately 9  nights (7.5 allocated nights   and 1.5
nights of  engineering time). Two  additional nights allocated  by the
SOAR partners for  the TESS follow-up were lost to  clouds. The seeing
in the  second half of 2023  has been consistently worse than  usual, often
precluding observations of faint targets. 


\subsection{Instrument, Observing Procedure, and Data Processing}
\label{sec:inst}

The   observations  reported   here  were   obtained  with   the  {\it
  high-resolution camera} (HRCam)  --- a fast imager  designed to work
at  the  4.1  m  SOAR telescope  \citep{HRCAM}.   The  instrument  and
observing  procedure are  described in  the previous  papers of  these
series \citep[e.g.][]{SAM19} and  briefly summarized by \citet{SAM22}.
There is  no need to  repeat the same text  here.  We used  mostly the
near-infrared  $I$ filter  (824/170\,nm),  while  the Str\"omgren  $y$
filter  (543/22\,nm)  was  used  for  brighter  and/or  closer  pairs.
Calibration of the pixel  scale and orientation is based on  a set of wide
binaries  with well-modeled  motion,  linked to  the Gaia  astrometry.
Typical external errors of positional measurements, evaluated from the
calibrators and from residuals to orbits,  are 1-2 mas; the errors are
larger for pairs  with large contrast or faint  magnitudes.  The total
number  of  HRCam  observations  made since  2008  approaches  40\,000
(26\,331 measurements and 12\,965 non-resolutions).

The diffraction limit  $\lambda/D$ of the 4.1 m SOAR  telescope in the
$y$ and $I$ filters is 27  and 41 mas, respectively. However, as shown
in  Figure~\ref{fig:dmsep},  many  pairs at  closer  separations  were
measured. Their positions  are obtained by modeling  the speckle power
spectra of  the target and of the reference star. Below  the diffraction
limit, only  part of the  elongated central fringe is  accessible, and
the  resulting positions  become  less accurate  (they  are marked  by
colons). Nevertheless,  the closest  pairs are  also the  fastest, and
even less accurate data are useful for tracing their orbital motion.

\section{Results}
\label{sec:res}

\subsection{Data Tables}
\label{sec:tables}

The results (measures of resolved pairs and non-resolutions) are presented in 
almost the same format as in \citet{SAM22}. The long tables are published
electronically; here we describe their content.


\begin{deluxetable}{ l l l l }
\tabletypesize{\scriptsize}
\tablewidth{0pt}
\tablecaption{Measurements of Double Stars at SOAR 
\label{tab:measures}}
\tablehead{
\colhead{Col.} &
\colhead{Label} &
\colhead{Format} &
\colhead{Description, units} 
}
\startdata
\phn1 & WDS                  & A10  & WDS code (J2000)  \\
\phn2 & Discov.\             & A16  & Discoverer Designation   \\
\phn3 & Other                & A16  & Alternative name \\
\phn4 & RA                   & F8.4 & R.A.\ J2000 (deg) \\
\phn5 & Dec                  & F8.4 & Declination J2000 (deg) \\
\phn6 & Epoch                & F9.4 & Julian year (yr) \\
\phn7 & Filt.\               & A2   & Filter \\
\phn8 & $N$                  & I2   & Number of averaged cubes \\
\phn9 & $\theta$             & F8.1 & Position angle (deg) \\
   10 & $\rho \sigma_\theta$ & F5.1 & Tangential error (mas) \\
   11 & $\rho$               & F8.4 & Separation (arcsec) \\
   12 & $\sigma_\rho$        & F5.1 & Radial error (mas) \\
   13 & $\Delta m$           & F7.1 & Magnitude difference (mag) \\
   14 & Flag                 & A1   & Flag of magnitude difference\tablenotemark{a} \\
   15 & Tag                 & A1   & System tag\tablenotemark{b} \\
   16 & (O$-$C)$_\theta$     & F8.1 & Residual in angle (deg) \\
   17 & (O$-$C)$_\rho$       & F8.3 & Residual in separation (arcsec) \\
   18 & Ref                  & A9   & Orbit reference\tablenotemark{c} 
\enddata
\tablenotetext{a}{Magnitude Flags: 
q -- the quadrant is determined; 
* -- $\Delta m$ and quadrant from average image; 
: -- noisy data or tentative measures. }
\tablenotetext{b}{System tags: 
A -- Hipparcos-Gaia acceleration stars (K.~Franson);
C -- Wide pairs observed for J.~Chanam\'e;
G -- Wide pairs with relative positions in Gaia DR3;
N -- New pair resolved in 2023;
P -- Hierarchical systems from \citet{Powell2023}; 
T -- Hierarchies within 100 pc \citep{GKM};
Z -- TESS objects of interest \citep{TESS2}.
}
\tablenotetext{c}{Orbit References are provided at
  \url{https://crf.usno.navy.mil/data_products/WDS/orb6/wdsref.html} }
\end{deluxetable}

\begin{deluxetable}{ l l l l }
\tabletypesize{\scriptsize}
\tablewidth{0pt}
\tablecaption{Unresolved Stars 
\label{tab:single}}
\tablehead{
\colhead{Col.} &
\colhead{Label} &
\colhead{Format} &
\colhead{Description, units} 
}
\startdata
\phn1 & WDS              & A10  & WDS code (J2000)  \\
\phn2 & Discov.\         & A16  & Discoverer Designation  \\
\phn3 & Other            & A16  & Alternative name \\
\phn4 & RA               & F8.4 & R.A. J2000 (deg) \\
\phn5 & Dec              & F8.4 & Declination J2000 (deg) \\
\phn6 & Epoch            & F9.4 & Julian year  (yr) \\
\phn7 & Filt.\           & A2   & Filter \\
\phn8 & $N$              & I2   & Number of averaged cubes \\
\phn9 & $\rho_{\rm min}$ & F7.3 & Angular resolution (arcsec)  \\
   10 & $\Delta m$(0.15) & F7.2 & Max. $\Delta m$ at 0\farcs15 (mag) \\
   11 & $\Delta m$(1)    & F7.2 & Max. $\Delta m$ at 1\arcsec (mag) \\
\enddata
\end{deluxetable}

A note  on system designations  is needed  here. The historic  data on
double  stars in  the  WDS  database \citep{WDS}  use  a tradition  of
assigning each pair  a unique Discoverer Designation  (DD), linking it
to the  first published  measurement.  For example,  COU~929 indicates
that this  pair has  been discovered visually  by P.~Couteau.   In the
modern  epoch,  most new  pairs  are  identified  in surveys  such  as
Hipparcos or Gaia;  WDS currently assigns them disparate  DDs based on
the first  author of  (usually multi-author) publications  listing the
pairs. Furthermore, new  subsystems in multiple stars  are often given
``recycled'' DDs of the wide pairs to which they belong, with modified
components qualifiers.  Nowadays the DDs  became obsolete and they are
no longer used  by the wider community.  They have  lost links both to
the primary data  source (e.g.  Gaia) and to the  original papers.  In
the                           WDS                          supplement,
\href{https://crf.usno.navy.mil/usnodocspath?file_name=WDS%20Supplemental%20Catalog%3A%20Summary&source_file=/data_products/WDS/Supplement/wdss_summ.txt}{WDSS},
  the DDs are no longer assigned.  The combination of the WDS code and
  component designations  like A,B  or Aa,Ab uniquely  identifies each
  pair, and the DDs are redundant.  Objects without existing DDs
  are identified in the data tables by the WDS codes, labels related to the
  observing  programs   (e.g.  GKM,  QUAD,  or   M25),  and  component
  designations if resolved.

Table~\ref{tab:measures} lists  1913 measures  of 1533  resolved pairs
and subsystems, including new discoveries. The pairs are identified by
their WDS-style  codes based on  the J2000 coordinates  and DDs
adopted   in  the  WDS  catalog   \citep{WDS}  or  their
substitutes, as  well as  by alternative names  in column  (3), mostly
from the Hipparcos catalog. Equatorial coordinates for the epoch J2000
in degrees  are given in  columns (4)  and (5) to  facilitate matching
with other  catalogs and  databases. Circumstances of  this particular
observation (Julian Year,  filter, number of  cubes), be it  Table 1 or  2, are
given in  columns (6) through  (8). In  the case of  resolved multiple
systems, the positional measurements  and their errors (columns 9--12)
and magnitude differences (column 13) refer to the individual pairings
between  components, not  to their  photocenters. As  in the  previous
papers of  this series, we list  the internal errors derived  from the
power  spectrum modeling  and from  the difference  between the  measures
obtained  from two  data cubes.  The real (external)  errors are  usually larger,
especially for difficult pairs with substantial $\Delta m$ and/or with
small  separations.

The flags in column (14) indicate the cases where the true quadrant is
determined (otherwise the position angle is measured modulo 180\degr),
when  the  relative photometry  of  wide  pairs  is derived  from  the
long-exposure  images  (this  reduces   the  bias  caused  by  speckle
anisoplanatism), and  when the data  are noisy or the  resolutions are
tentative (see TMH10).

To facilitate identification  of pairs that either  have been resolved
previously but remain unpublished or are published but not yet entered
in the WDS, we provide in column (15) one-character tags as follows: A
--- acceleration stars (Franson, Bowler), C -- wide pairs (Chanam\'e),
G ---  wide pairs  with relative  positions in GDR3,  N ---  new pairs
resolved  here  for the  first  time  (Section \ref{sec:new}),  P  ---
hierarchical systems published by \citet{Powell2023}, T --- components
of triple  or higher-order  hierarchies within 100\,pc  \citep{GKM}, Z
--- TESS  objects of  interest  \citep{TESS2}.  For  those pairs,  the
column (2)  contains program-related labels and  component's pairings,
e.g. GKM~Aa,Ab, instead of DDs.

For binary stars with known orbits,  the residuals to the latest orbit
and its reference are provided  in columns (16)--(18). Residuals close
to 180\degr  ~mean that the orbit  swaps the brighter (A)  and fainter
(B)  stars. However,  in some  binaries  or triples  the secondary  is
fainter in one filter and brighter in the other. In these cases, it is
better  to keep  the historical  identification of  the components  in
agreement  with  the  orbit  and   to  provide  a  negative  magnitude
difference $\Delta m$.

The  non-resolutions  of  552  targets (mostly  reference  stars)  are
reported in Table~\ref{tab:single}.  Its first columns (1) to (8) have
the same meaning and format as in Table~\ref{tab:measures}. Column (9)
gives the minimum resolvable separation when pairs with $\Delta m < 1$
mag are detectable. It is  computed from the maximum spatial frequency
of the  useful signal in the  power spectrum and is  normally close to
the  formal  diffraction  limit $\lambda/D$.  The  following
columns (10) and (11) provide  the indicative dynamic range, i.e., the
maximum magnitude difference at separations of 0\farcs15 and 1\arcsec,
respectively, at $5\sigma$ detection level. 


\subsection{New Pairs}
\label{sec:new}

Newly    resolved   pairs    are   marked    by   the    tag   N    in
Table~\ref{tab:measures}, so there  is no need to  give their separate
list. There are  61 measures with this tag referring  to 38 pairs (some
of them have multiple measures). This does not include the wider pairs
with tag G present  in the GDR3 but not yet  recognized as such in
the WDS.  Comments  on 11 new subsystems in  nearby hierarchies (label
GKM) are given below in Section \ref{sec:GKM}. Four new pairs from the TESS
program have tags Z, so the total number of first-time resolutions is
42. 

Two  subsystems  in  nearby  triple M-type  dwarfs,  18113$-$7859  and
20253$-$2259, have been discovered in 2019.5 and 2022.4, respectively.
Their measures should have been published by \citet{Vrijmoet2022}, but
were omitted  owing to  a technical oversight,  while the  outer pairs
were reported.  All  measures of these subsystems  are published here.
The first pair has  an orbit with a period of 1.7  yr that will appear
in the forthcoming paper by Vrijmoet et al.

Of the 22  observed doubly eclipsing quadruples (label  QUAD) from the
paper by \citet{Kostov2022}, 10 were  resolved. These stars are mostly
faint, early-type, and distant. The  resolved objects are entered in the
MSC  as  2+2  quadruples,  but   in  fact  they  may  be  higher-order
hierarchies if the  resolved secondaries are not  eclipsing, while the
doubly eclipsing objects belong to the brighter primaries.

As noted above, we observed 15 candidate faint M dwarf binaries within
25\,pc (label M25). Two were resolved. The resolution of 01465$-$5340 at
0\farcs03  is  tentative  (it  has  been  confirmed  in  2024),  while
03287$-$1537 is resolved securely at 0\farcs33.

As in  the previous years,  we discovered  that a few  Hipparcos stars
observed  as point-source  references had  close and  faint companions
(HIP 8387, 9534, 41375, 47070, 72640).  One of those, HIP~47070, also has a
physical companion BD$-18^\circ$~2729 at  51\farcs4. The estimated period of
the inner 0\farcs23  pair Aa,Ab is 60 yr, and  its GDR3 astrometry does
not show (yet)  deviations from the linear motion.  Other resolved Hipparcos
stars were targeted on purpose because they had strong indications of
binarity in the Gaia data (HIP 16526, 37505, 42728, 96562).

{\em 01350$-$0430  Aa,Ab} is the  brighter ($V=14.98$ mag)  component of
the 40\arcsec ~nearby pair LDS 5338 AB  at 80\,pc from the Sun. It was
observed on  suggestion by D.~Nazor  who noted that GDR3 contains
two equal stars  at 0\farcs35 separation.  Star A  was indeed resolved
at 0\farcs307. This triple is added to the MSC.  The distant companion
LDS 5338 B is a white dwarf.

The  star BD$-$17$^\circ$~1383  (06100$-$1702)  has  been observed  as
potential target  for long-baseline interferometry; its  resolution at
0\farcs028  is  below  the  estimated detection  limit  and  might  be
spurious. Another interferometric  candidate 18042$-$6430 was resolved
at 0\farcs037 in 2023.325, but found single in 2023.663; its estimated
period is 1 yr.

{\em 12592$-$6256  Aa,Ab (HIP 63377)}  has been resolved by  the speckle
camera   at  the   Gemini-S   telescope   (R.~Mendez,  2023,   private
communication) and  is confirmed  here, while  the wider  companion at
0\farcs45 (TOK~422) has been discovered at SOAR in 2016. The estimated
period of  the inner  pair is  3\,yr, double lines  were noted  in its
spectrum,  although GDR3 does  not give  yet its  orbit. This  is
another solar-type triple within 100\,pc. 

{\em  21199$-$5327 Ba,Bb}  is the  secondary star  in the  bright binary
$\theta$~Indi  (HJ   5258  AB,  7\farcs3).   H.~Zirm   (2023,  private
communication) has  computed astrometric orbit of  the subsystem Ba,Bb
with a period of  26 yr and suggested that it  should be resolvable by
speckle interferometry.  His prediction  has been confirmed in 2023.32
and 2023.57. The  magnitude difference of Ba,Bb is 5.4  mag in $y$ and
3.7 mag  in $I$,  and the  masses of  Ba and  Bb estimated  from their
absolute magnitudes  are 1.08 and 0.54  \msun, respectively. According
to the  Zirm's orbit, Ba,Bb is  closing down and it  will pass through
the periastron  in  2027.3.   The  brighter star  A  has  been  reportedly
resolved  in   2012.56  at   62\,mas  and  designated   as  MRN~3Aa,Ab
\citep{Marion2014}.   However, this  resolution,  never confirmed,  is
likely spurious,  considering the constant  radial velocity of  star A
\citep{Lagrange2009} and its good Gaia astrometry.


\subsection{New and Updated Orbits}
\label{sec:orbits}

\begin{figure*}[ht]
\epsscale{1.1}
\plotone{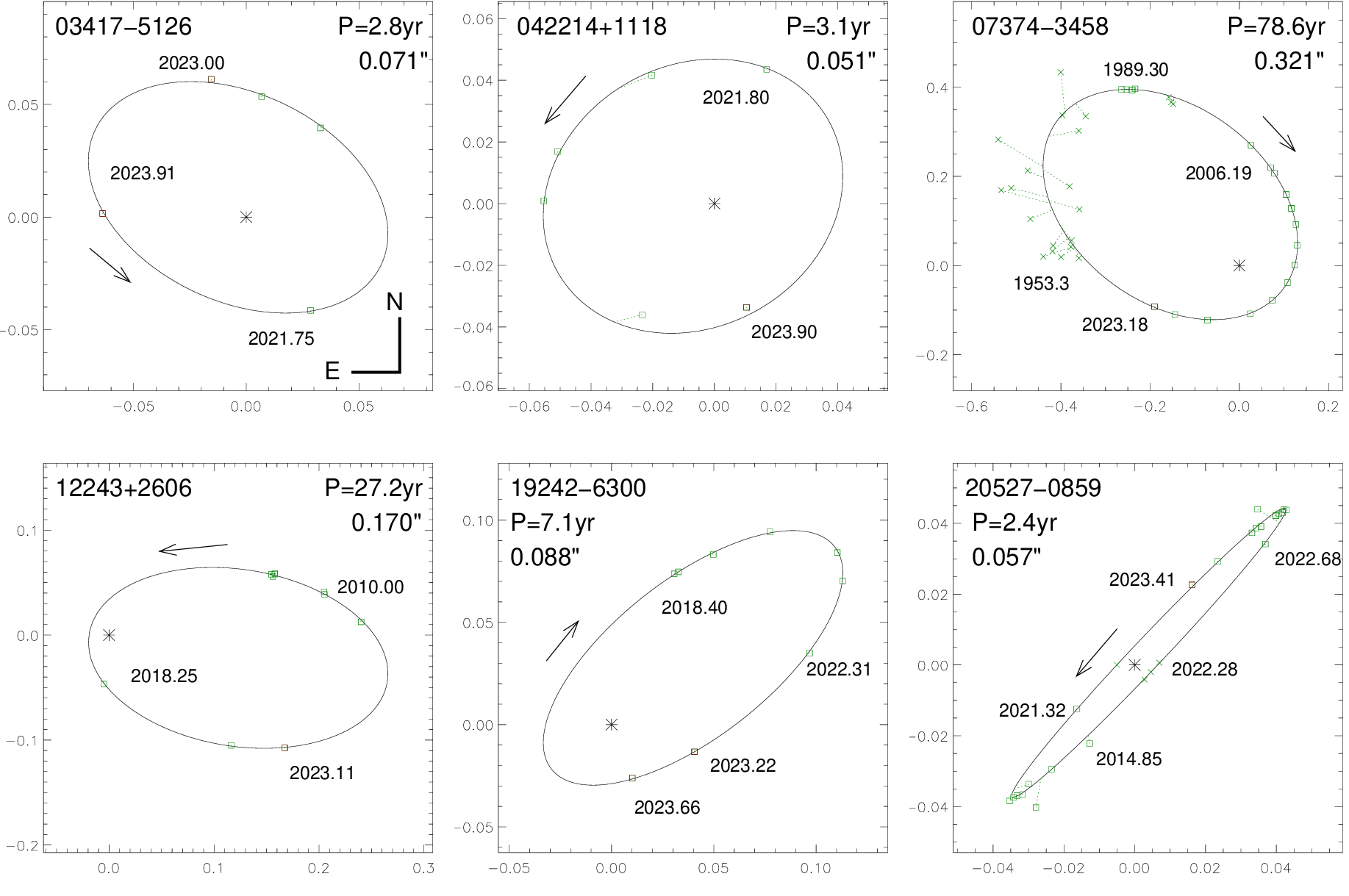}
\caption{Six new visual orbits with well-constrained orbital elements.
  In each plot, the ellipse  shows the fitted orbit, squares connected
  to the ellipse are accurate  speckle data, while visual measures and
  non-resolutions  are  shown  as   crosses.   The  primary  component
  (asterisk)  is   at  coordinate  origin,   the  axis  scale   is  in
  arcseconds. Orientation,  periods,  semimajor  axes,  and  sense  of
  rotation are indicated.
\label{fig:visual} }
\end{figure*}

The  seven Campbell  orbital  elements were  fitted  to all  available
positional measurements using the  IDL code {\tt orbit} \citep{orbit}.
The weights are inversely proportional  to the squares of the position
errors, assumed  to be 2\,mas  for the  SOAR data (larger  for fainter
stars or  pairs with large  contrast), 5\,mas for other  speckle data,
and 0\farcs05  or larger for visual  measurements.  The $\chi^2/(N-M)$
goodness-of-fit parameter is typically on the order of one, indicating
that the adopted errors are  approximately correct.  When the orbit is
poorly constrained, some elements are  fixed to values that match both
the positions and the expected mass sum, computed with parallaxes from
GDR3 or other sources.  The  longitude of periastron $\omega$ is fixed
to zero in  the degenerate cases of circular or  face-on orbits.  Note
that in our  previous paper \citep{SAM22}, the  orbits were calculated
using a different tool and a different weighting scheme.

The list of  147 computed orbits is divided into  two groups, reliable
(87) and  preliminary (60), using criteria  outlined by \citet{SAM22}.
In these  Tables \ref{tab:vborb}  and \ref{tab:vborb4}, the  system is
identified by  the WDS J2000  coordinate and the DD  (when available),
followed by  the seven Campbell  elements.  The two  rightmost columns
contain the  orbit grade  and a  reference to  the most  current orbit
which has  been here improved,  when available; the  first-time orbits
are referenced  as SOAR2023. The  grades are assigned by  the standard
method described by \citet{VB6}, which  includes both the rms residual
as  well as  the phase  or position  angle coverage,  total number  of
measures, and  the total number of  orbits.  The grades do  not always
correlate  with the  quality  metrics based  strictly  on absolute  or
relative errors of the elements  determined by the least-squares fits,
so  that  some orbits  in  Table~\ref{tab:vborb}  would be  considered
preliminary by their grades of  4 and 5.  In Table~\ref{tab:vborb} the
errors of each orbital element are listed in the subsequent line below
the main  entry, while  in Table~\ref{tab:vborb4}  the errors  are not
provided.

Many  orbits  in  the  present  lists are  based  exclusively  on  the
positions measured  at SOAR. It would  be desirable to obtain  data on
these  pairs  from   other  teams  for  a   cross-check.   Some  pairs
(04237+1131  and others)  are Hipparcos  ``problem'' stars  (suspected
non-single, aceleration, or stochastic solutions).

Figure~\ref{fig:visual}  contains  illustrative  plots of  six  visual
orbits computed here for the first time (except one), but nevertheless
well  constrained.  The  pair 03417$-$5126  is a  Hipparcos star  with
acceleration, first resolved  at SOAR in 2021 (tag A,  to be published
by  Franson  and Bowler);  its  preliminary  orbit  is based  on  five
measures  covering more  than  half  of the  2.8  yr period.   Regular
monitoring of  04224+1118, a neglected  binary in the  Hyades (PAT~5),
enables calculation of the first orbit. The orbit of 07374$-$3458 (FIN
324AC)  is  an update  of  the  previous solution  \citep{TMH15}.   It
highlights the vastly  improved accuracy of the  speckle data compared
to the historic visual measurements.  The  SOAR data from 2008 to 2023
cover the periastron, and no substantial orbit improvement is expected
in the coming decade.  Note that  the pair FIN~324AB listed in the WDS
is spurious. If it were real, it would make the triple system unstable
and would certainly  affect the observed motion of AC,  which shows no
deviations  from the  orbit  apart from  the  measurement errors  (the
weighted residuals are 2\,mas). The  pair 12243+2606 has been resolved
by the  Yale speckle program (YSC  97), but most data  come from SOAR;
the  measures   of  19242$-$6300  (TOK   799)  come  only   from  this
program. Finally, the orbit of  the bright speckle binary 20527$-$0859
(McA  64  Aa,Ab)  has  not been  determined  previously  despite  many
measures because of the short period (2.4 yr) and the lack of frequent
visits. The  puzzle has been  eventually solved  with the help  of the
dense coverage at  SOAR, including the times  of non-resolutions; four
positions corresponding  to non-resolutions are plotted  by crosses in
Figure~\ref{fig:visual} only  for illustration.   The quality  of this
orbit  is  excellent, with  weighted  residuals  of 1\,mas  (some  old
speckle data were given reduced  weights).  Yet, the official grade is
only 4,  illustrating a limitation  of the current grading  system and
possibly  an over-reliance  on the  total number  of measures  and the
total phase coverage.  The 4.3 yr  spectroscopic orbit of this star by
\citet{Abt1985}  is spurious;  a plot  of  their RVs  phased with  the
correct period reveals only noise.

Below we comment on some other pairs with new orbits.

{\it 05321$-$0305 (V1311 Ori)} is a system of pre-main sequence M-type
dwarfs containing at least six stars \citep{V1311Ori}. The updated 150
yr orbit of Aa,Ab uses measures near the periastron recently published
by \citet{Calissendorff2022}  and the  latest SOAR  data, but  it will
remain loosely constrained until the binary approaches the apastron in
a few  decades.  On the  other hand, the tentative  8 yr orbit  of the
faster pair Ba,Bb discovered at  SOAR in 2021 will become well-defined
in the next few years.

{\it  09525$-$0806  (HR  3909).   }    The  very  bright  double  star
$\gamma$~Sex was first split by Alvan  Clark in 1854 while testing his
telescopes \citep{Clark1857}.   The first resolution of  this pair and
others, and the  praise given to his telescopes by  W.~R.~Dawes led to
Clark and his sons getting many orders for telescopes and constructing
the largest telescope in the world six different times in the 19th and
early 20th centuries. The first  accurate measure would wait more than
two decades until  \citet{Hall1892} used a larger  Clark refractor. We
update the previous  77.8 yr orbit by  \citet{USN2007a} using accurate
speckle measures; its further improvement is foreseen as the pair goes
through the periastron in 2034.

{\it 13175$+$2024 (YSC 149)} is a resolved triple system where the inner
pair Aa,Ab has a period around 5 yr. Pending the full analysis, a
preliminary 33 yr outer orbit is given here in Table~\ref{tab:vborb4}.

{\it 16090$-$0939 (HIP 79122)} is a $\gamma$ Dor variable of spectral
type A8V, originally observed at  the request of Frank  Fekel. The
composite spectrum  revealed evidence of two  stars \citep{Henry2011}.
This binary was first measured in 2008 with an early version of HRCam,
when  the USNO  speckle camera  did  not arrive  at CTIO  in time  for
observing on the Blanco telescope.  The pair with $\Delta m \sim 3.5$
mag  is not  resolved at  close separations  near the  periastron, the
eccentricity is  large but not well  constrained, so it was  fixed at
$e=0.97$ to obtain the mass sum around 5 \msun with the GDR3 parallax
of 12.57$\pm$0.09\,mas.

\startlongtable

\begin{deluxetable*}{l l cccc ccc cc}    
\tabletypesize{\scriptsize}     
\tablecaption{Visual Orbits with well-defined Errors
\label{tab:vborb}          }
\tablewidth{0pt}                                   
\tablehead{                                                                     
\colhead{WDS} & 
\colhead{Discoverer} & 
\colhead{$P$} & 
\colhead{$T$} & 
\colhead{$e$} & 
\colhead{$a$} & 
\colhead{$\Omega$ } & 
\colhead{$\omega$ } & 
\colhead{$i$ } & 
\colhead{Grade }  &
\colhead{Ref.} \\
\colhead{$\alpha$, $\delta$ (2000)} &
\colhead{Designation} &  
\colhead{(yr)} &
\colhead{(yr)} & &
\colhead{(arcsec)} & 
\colhead{(deg)} & 
\colhead{(deg)} & 
\colhead{(deg)} &  & 
}
\startdata
00026$+$1841 & HDS 2 Aa,Ab & 22.68 & 2020.967 & 0.631 & 0.1106 & 17.4 & 302.2 & 59.8 & 3 & SOAR2023 \\
             &     & $\pm$0.34 & $\pm$0.074 & $\pm$0.013 & $\pm$0.0028 & $\pm$2.3 & $\pm$3.1 & $\pm$1.3&     &  \\
00061$+$0943 & HDS 7 & 51.23 & 2003.83 & 0.534 & 0.2145 & 184.1 & 237.1 & 60.0 & 3 & FRM2017c \\
             &     & $\pm$1.11 & $\pm$0.11 & $\pm$0.014 & $\pm$0.0032 & $\pm$1.1 & $\pm$2.1 & $\pm$0.9&     &  \\
00098$-$3347 & SEE 3 & 295.7 & 1978.66 & 0.770 & 0.928 & 274.3 & 72.5 & 33.1 & 4 & Hrt2010a \\
             &     & $\pm$8.6 & $\pm$0.48 & $\pm$0.009 & $\pm$0.020 & $\pm$3.4 & $\pm$3.4 & $\pm$2.0&     &  \\
00111$+$0513 & TOK 869 & 8.49 & 2018.22 & 0.183 & 0.1112 & 108.8 & 239.7 & 51.2 & 4 & Tok2023a \\
             &     & $\pm$0.10 & $\pm$0.18 & $\pm$0.018 & $\pm$0.0029 & $\pm$2.0 & $\pm$8.0 & $\pm$1.8&     &  \\
00135$-$3650 & HDS 32 & 15.272 & 2009.080 & 0.242 & 0.2233 & 101.2 & 266.0 & 159.6 & 2 & Tok2017b \\
             &     & $\pm$0.054 & $\pm$0.078 & $\pm$0.005 & $\pm$0.0016 & $\pm$3.8 & $\pm$3.4 & $\pm$1.8&     &  \\
00143$-$2732 & HDS 33 & 10.150 & 2013.465 & 0.608 & 0.1222 & 33.3 & 82.5 & 27.6 & 2 & Tok2015c \\
             &     & $\pm$0.016 & $\pm$0.017 & $\pm$0.004 & $\pm$0.0010 & $\pm$2.7 & $\pm$2.4 & $\pm$1.1&     &  \\
01017$+$2518 & HDS 134 & 16.563 & 2015.209 & 0.396 & 0.1303 & 237.1 & 288.0 & 59.0 & 2 & AlW2014 \\
             &     & $\pm$0.056 & $\pm$0.070 & $\pm$0.012 & $\pm$0.0017 & $\pm$0.7 & $\pm$0.6 & $\pm$0.7&     &  \\
01150$-$0908 & UC 527  Aa,Ab & 3.09 & 2022.40 & 0.40 & 0.0407 & 110.5 & 260.2 & 51.5 & 4 & SOAR2023 \\
             &     & $\pm$0.21 & $\pm$0.09 & $\pm$0.05 & $\pm$0.0029 & $\pm$9.5 & $\pm$7.0 & $\pm$6.0&     &  \\
01158$-$6853 & I 27 CD & 85.14 & 2001.25 & 0.040 & 1.0877 & 140.6 & 132.6 & 31.1 & 3 & Izm2019 \\
             &     & $\pm$0.21 & $\pm$0.92 & $\pm$0.002 & $\pm$0.0054 & $\pm$0.9 & $\pm$3.3 & $\pm$0.3&     &  \\
01406$+$0846 & TOK 872 & 5.979 & 2004.600 & 0.710 & 0.0770 & 122.9 & 245.6 & 60.3 & 3 & SOAR2023 \\
             &     & $\pm$0.026 & $\pm$0.004 & $\pm$0.006 & $\pm$0.0025 & $\pm$2.6 & $\pm$0.8 & $\pm$2.0&     &  \\
01497$-$1022 & BU 1168 & 121.0 & 1924.56 & 0.700 & 0.2949 & 205.5 & 106.8 & 94.9 & 4 & SOAR2023 \\
             &     & $\pm$3.5 & $\pm$3.48 & fixed & $\pm$0.0119 & $\pm$0.7 & $\pm$2.5 & $\pm$0.6&     &  \\
02166$-$5026 & TOK 185 & 10.572 & 2012.51 & 0.023 & 0.0907 & 266.6 & 338.9 & 47.1 & 2 & Tok2022f \\
             &     & $\pm$0.062 & $\pm$0.80 & $\pm$0.007 & $\pm$0.0025 & $\pm$2.2 & $\pm$27.3 & $\pm$1.4&     &  \\
02167$+$0632 & YSC 20 & 13.291 & 2010.630 & 0.418 & 0.1254 & 45.9 & 31.2 & 27.4 & 3 & Hor2021 \\
             &     & $\pm$0.045 & $\pm$0.053 & $\pm$0.008 & $\pm$0.0021 & $\pm$6.6 & $\pm$7.0 & $\pm$3.8&     &  \\
02405$-$2408 & SEE 19 & 307.2 & 2016.76 & 0.832 & 0.379 & 235.5 & 81.6 & 148.0 & 3 & Tok2015c \\
             &     & $\pm$24.4 & $\pm$0.05 & $\pm$0.009 & $\pm$0.020 & $\pm$2.7 & $\pm$3.1 & $\pm$1.4&     &  \\
02415$-$1506 & HDS 351 & 42.7 & 1979.8 & 0.374 & 0.241 & 24.2 & 294.6 & 88.8 & 3 & Tok2022f \\
             &     & $\pm$6.0 & $\pm$6.5 & $\pm$0.041 & $\pm$0.025 & $\pm$0.5 & $\pm$17.0 & $\pm$0.4&     &  \\
03309$-$6200 & TOK 190 & 3.627 & 2021.081 & 0.445 & 0.0637 & 30.4 & 242.5 & 79.2 & 2 & Tok2018i \\
             &     & $\pm$0.011 & $\pm$0.027 & $\pm$0.023 & $\pm$0.0011 & $\pm$0.8 & $\pm$1.8 & $\pm$0.8&     &  \\
04224$+$1118 & PAT 5 & 3.049 & 2024.18 & 0.162 & 0.0508 & 121.4 & 124.6 & 32.9 & 2 & SOAR2023 \\
             &     & $\pm$0.014 & $\pm$0.101 & $\pm$0.032 & $\pm$0.0020 & $\pm$9.5 & $\pm$11.3 & $\pm$9.4&     &  \\
04279$+$2427 & TOK 877 & 10.9 & 2015.15 & 0.192 & 0.156 & 85.9 & 221.1 & 63.1 & 4 & SOAR2023 \\
             &     & $\pm$1.0 & $\pm$0.25 & $\pm$0.089 & $\pm$0.012 & $\pm$3.4 & $\pm$10.4 & $\pm$3.1&     &  \\
04506$+$1505 & CHR 20 & 5.738 & 2003.825 & 0.048 & 0.0890 & 129.7 & 270.5 & 113.2 & 2 & Doc2018h \\
             &     & $\pm$0.006 & $\pm$0.085 & $\pm$0.010 & $\pm$0.0010 & $\pm$0.7 & $\pm$5.2 & $\pm$0.6&     &  \\
05301$-$3228 & B 1946 & 67.2 & 2023.24 & 0.776 & 0.1140 & 180.2 & 358.7 & 34.9 & 3 & Tok2022f \\
             &     & $\pm$4.0 & $\pm$0.07 & $\pm$0.008 & $\pm$0.0014 & $\pm$5.5 & $\pm$7.5 & $\pm$4.2&     &  \\
05427$-$6708 & I 745 & 205.1 & 2017.70 & 0.786 & 0.542 & 244.9 & 205.7 & 75.4 & 4 & Tok2019c \\
             &     & $\pm$35.1 & $\pm$0.18 & $\pm$0.024 & $\pm$0.061 & $\pm$0.8 & $\pm$1.5 & $\pm$0.7&     &  \\
06460$-$6624 & TOK 826 CD & 11.70 & 2023.000 & 0.621 & 0.0970 & 36.1 & 319.5 & 149.1 & 4 & SOAR2023 \\
             &     & $\pm$1.86 & $\pm$0.048 & $\pm$0.044 & $\pm$0.0051 & $\pm$11.6 & $\pm$14.5 & $\pm$8.9&     &  \\
06481$-$0948 & A 1056 & 113.6 & 1995.91 & 0.712 & 0.405 & 70.4 & 257.8 & 78.5 & 3 & Tok2015c \\
             &     & $\pm$3.7 & $\pm$0.65 & $\pm$0.047 & $\pm$0.022 & $\pm$1.5 & $\pm$1.5 & fixed&     &  \\
07374$-$3458 & FIN 324 AC & 78.58 & 2016.772 & 0.656 & 0.3208 & 71.8 & 205.7 & 153.3 & 2 & Tok2015c \\
             &     & $\pm$0.73 & $\pm$0.021 & $\pm$0.002 & $\pm$0.0023 & $\pm$1.7 & $\pm$1.6 & $\pm$0.7&     &  \\
08250$-$4246 & CHR 226 Ba,Bb & 63.2 & 2018.04 & 0.502 & 0.0655 & 82.1 & 259.5 & 114.7 & 3 & Tok2023d \\
             &     & $\pm$7.7 & $\pm$0.35 & $\pm$0.035 & $\pm$0.0054 & $\pm$2.0 & $\pm$7.8 & $\pm$2.5&     &  \\
08277$-$0425 & A 550 & 21.062 & 2002.583 & 0.844 & 0.0939 & 34.5 & 47.8 & 150.5 & 2 & Tok2023a \\
             &     & $\pm$0.051 & $\pm$0.069 & $\pm$0.009 & $\pm$0.0027 & $\pm$8.2 & $\pm$8.3 & $\pm$3.8&     &  \\
08313$-$0601 & BAG 49 Aa,Ab & 16.43 & 2007.60 & 0.718 & 0.247 & 122.0 & 253.6 & 72.7 & 3 & SOAR2023 \\
             &     & $\pm$0.13 & $\pm$0.14 & $\pm$0.046 & $\pm$0.021 & $\pm$2.3 & $\pm$1.8 & $\pm$1.9&     &  \\
08375$-$5336 & GKM  Aa,Ab & 2.90 & 2023.13 & 0.496 & 0.0520 & 66.4 & 199.9 & 41.7 & 3 & SOAR2023 \\
             &     & $\pm$0.22 & $\pm$0.005 & $\pm$0.048 & $\pm$0.0025 & $\pm$7.3 & $\pm$10.9 & $\pm$8.7&     &  \\
09525$-$0806 & AC 5 AB & 77.61 & 1956.84 & 0.741 & 0.380 & 198.5 & 306.2 & 143.2 & 2 & USN2007a \\
             &     & $\pm$0.59 & $\pm$0.69 & $\pm$0.019 & $\pm$0.013 & $\pm$5.2 & $\pm$6.5 & $\pm$3.4&     &  \\
10255$-$6504 & HDS 1499 & 20.44 & 2022.053 & 0.782 & 0.0815 & 130.1 & 132.9 & 36.4 & 3 & Tok2023a \\
             &     & $\pm$0.73 & $\pm$0.055 & $\pm$0.010 & $\pm$0.0019 & $\pm$4.7 & $\pm$5.4 & $\pm$3.8&     &  \\
11238$-$3829 & CHR 241 & 3.530 & 2013.182 & 0.380 & 0.0582 & 155.7 & 254.8 & 111.8 & 2 & Tok2016e \\
             &     & $\pm$0.009 & $\pm$0.031 & $\pm$0.019 & $\pm$0.0009 & $\pm$1.0 & $\pm$1.4 & $\pm$0.9&     &  \\
12018$-$3439 & I 215 AB & 178.48 & 2016.29 & 0.407 & 0.862 & 269.8 & 268.4 & 111.3 & 3 & Tok2015a \\
             &     & $\pm$4.50 & $\pm$0.43 & $\pm$0.006 & $\pm$0.015 & $\pm$0.3 & $\pm$2.3 & $\pm$0.3&     &  \\
12104$-$4352 & TOK 897 & 12.76 & 2017.90 & 0.597 & 0.0630 & 86.1 & 101.8 & 50.2 & 3 & Tok2022f \\
             &     & $\pm$0.69 & $\pm$0.05 & $\pm$0.023 & $\pm$0.0025 & $\pm$3.3 & $\pm$3.4 & $\pm$2.0&     &  \\
12114$-$1647 & S 634 Aa,Ab & 0.5792 & 2018.7250 & 0.2809 & 0.0253 & 49.3 & 284.2 & 43.8 & 2 & SOAR2023 \\
             &     & $\pm$0.0006 & $\pm$0.0007 & $\pm$0.0018 & $\pm$0.0021 & $\pm$4.2 & $\pm$0.5 & $\pm$8.6&     &  \\
12243$+$2606 & YSC 97 & 27.25 & 2017.80 & 0.868 & 0.170 & 164.0 & 275.1 & 32.3 & 4 & Tok2023d \\
             &     & $\pm$1.76 & $\pm$0.11 & $\pm$0.021 & $\pm$0.012 & $\pm$8.2 & $\pm$7.1 & $\pm$7.3&     &  \\
12386$-$2704 & BWL 32 & 17.85 & 2019.996 & 0.334 & 0.2296 & 188.5 & 74.3 & 135.2 & 3 & Tok2022f \\
             &     & $\pm$0.65 & $\pm$0.093 & $\pm$0.016 & $\pm$0.0046 & $\pm$3.1 & $\pm$5.7 & $\pm$1.6&     &  \\
12455$-$4552 & HDS 1789 Aa,Ab & 23.41 & 2024.00 & 0.248 & 0.127 & 33.8 & 357.7 & 33.2 & 3 & Tok2023d \\
             &     & $\pm$0.45 & $\pm$0.38 & $\pm$0.059 & $\pm$0.005 & $\pm$13.5 & $\pm$21.4 & $\pm$5.4&     &  \\
12479$-$5127 & TOK 720 & 10.78 & 2023.77 & 0.292 & 0.0505 & 85.7 & 134.1 & 152.2 & 3 & Tok2022f \\
             &     & $\pm$1.14 & $\pm$0.18 & $\pm$0.084 & $\pm$0.0045 & $\pm$16.1 & $\pm$15.1 & $\pm$11.1&     &  \\
12528$+$1225 & TOK 401 & 9.964 & 2015.56 & 0.097 & 0.1108 & 117.3 & 294.5 & 75.1 & 3 & Tok2018b \\
             &     & $\pm$0.080 & $\pm$0.13 & $\pm$0.011 & $\pm$0.0010 & $\pm$0.6 & $\pm$5.0 & $\pm$0.6&     &  \\
13132$-$0501 & TOK 402 & 17.48 & 2016.694 & 0.583 & 0.1534 & 114.7 & 254.4 & 111.2 & 3 & Tok2020g \\
             &     & $\pm$0.30 & $\pm$0.016 & $\pm$0.010 & $\pm$0.0014 & $\pm$0.4 & $\pm$0.7 & $\pm$0.5&     &  \\
13133$+$1621 & DOC 1 & 25.44 & 2014.81 & 0.424 & 0.0852 & 95.0 & 18.6 & 61.0 & 4 & Doc2018h \\
             &     & $\pm$0.17 & $\pm$0.16 & $\pm$0.013 & $\pm$0.0018 & $\pm$2.1 & $\pm$4.2 & $\pm$3.6&     &  \\
13513$-$2423 & WSI 77 & 10.490 & 2009.258 & 0.342 & 0.2848 & 171.6 & 140.2 & 96.4 & 2 & Tok2012b \\
             &     & $\pm$0.006 & $\pm$0.010 & $\pm$0.002 & $\pm$0.0006 & $\pm$0.1 & $\pm$0.4 & $\pm$0.1&     &  \\
13598$-$0333 & HDS 1962 & 9.776 & 2008.342 & 0.396 & 0.0791 & 37.0 & 233.1 & 56.7 & 2 & Tok2019c \\
             &     & $\pm$0.056 & $\pm$0.057 & $\pm$0.011 & $\pm$0.0010 & $\pm$1.2 & $\pm$1.7 & $\pm$0.8&     &  \\
14219$-$3126 & BWL 38 & 9.77 & 2020.221 & 0.371 & 0.0997 & 236.8 & 287.0 & 106.0 & 3 & Tok2022f \\
             &     & $\pm$0.15 & $\pm$0.093 & $\pm$0.035 & $\pm$0.0029 & $\pm$1.9 & $\pm$4.5 & $\pm$2.3&     &  \\
14453$-$3609 & I 528 AB & 15.95 & 2023.47 & 0.628 & 0.0484 & 226.5 & 76.1 & 32.8 & 2 & Tok2022a \\
             &     & $\pm$0.17 & $\pm$0.11 & $\pm$0.045 & $\pm$0.0038 & $\pm$12.2 & $\pm$10.5 & $\pm$9.2&     &  \\
14509$-$1603 & BEU 19 Ba,Bb & 16.229 & 2021.95 & 0.244 & 0.346 & 122.7 & 184.1 & 54.4 & 4 & Tok2023d \\
             &     & $\pm$0.043 & $\pm$0.12 & $\pm$0.008 & $\pm$0.006 & $\pm$1.4 & $\pm$3.3 & $\pm$1.5&     &  \\
15006$+$0836 & YSC 8 AB & 6.922 & 2016.878 & 0.375 & 0.1163 & 149.4 & 99.7 & 95.9 & 2 & Tok2018 \\
             &     & $\pm$0.003 & $\pm$0.007 & $\pm$0.002 & $\pm$0.0006 & $\pm$0.3 & $\pm$0.4 & $\pm$0.3&     &  \\
15071$-$0217 & A 689 & 67.26 & 1994.46 & 0.671 & 0.2194 & 145.1 & 12.6 & 107.4 & 5 & Doc2023 \\ 
             &     & $\pm$0.92 & $\pm$0.41 & $\pm$0.015 & $\pm$0.0032 & $\pm$0.8 & $\pm$3.6 & $\pm$1.2&     &  \\
15332$-$2429 & SEE 238 Ba,Bb & 61.70 & 1999.82 & 0.676 & 0.228 & 204.7 & 132.5 & 24.0 & 2 & Msn2017a \\
             &     & $\pm$0.41 & $\pm$0.30 & $\pm$0.010 & $\pm$0.007 & $\pm$7.2 & $\pm$7.8 & $\pm$5.5&     &  \\
15367$-$4208 & TOK 408 Ca,Cb & 7.97 & 2014.41 & 0.0 & 0.0570 & 102.0 & 0.0 & 62.9 & 4 & Tok2018b \\
             &     & $\pm$0.16 & $\pm$0.15 & fixed & $\pm$0.0015 & $\pm$3.3 & fixed & $\pm$5.1&     &  \\
15481$-$5811 & SKF 2839 Aa,Ab & 11.11 & 2023.771 & 0.565 & 0.1704 & 14.3 & 46.3 & 160.5 & 3 & SOAR2023 \\
             &     & $\pm$0.45 & $\pm$0.018 & $\pm$0.015 & $\pm$0.0020 & $\pm$10.6 & $\pm$8.4 & $\pm$6.8&     &  \\
16054$-$1948 & BU 947 AB & 220.0 & 2054.5 & 0.685 & 0.851 & 80.2 & 242.2 & 75.7 & 3 & Tok2017c \\
             &     & fixed & $\pm$3.5 & $\pm$0.021 & $\pm$0.032 & $\pm$1.5 & $\pm$1.8 & $\pm$0.9&     &  \\
16054$-$1948 & MCA 42 CE & 18.953 & 2006.00 & 0.610 & 0.1047 & 108.0 & 80.8 & 45.4 & 2 & Tok2018e \\
             &     & $\pm$0.064 & $\pm$0.12 & $\pm$0.010 & $\pm$0.0016 & $\pm$2.9 & $\pm$1.7 & $\pm$1.3&     &  \\
16077$-$2125 & MCA 43 & 11.32 & 2017.06 & 0.40 & 0.0476 & 143.0 & 199.9 & 81.6 & 5 & SOAR2023 \\ 
             &     & $\pm$0.11 & $\pm$0.68 & $\pm$0.11 & $\pm$0.0030 & $\pm$1.7 & $\pm$17.4 & $\pm$2.3&     &  \\
16090$-$0939 & WSI 85 & 9.51 & 2020.89 & 0.970 & 0.098 & 139.6 & 319.6 & 83.5 & 3 & SOAR2023 \\
             &     & $\pm$0.15 & $\pm$0.31 & fixed & $\pm$0.028 & $\pm$3.5 & $\pm$20.5 & $\pm$5.6&     &  \\
16249$-$2240 & KSA 129 Aa,Ab & 13.13 & 2019.45 & 0.290 & 0.0584 & 67.5 & 143.0 & 165.2 & 3 & Tok2023d \\
             &     & $\pm$0.34 & $\pm$0.37 & $\pm$0.048 & $\pm$0.0087 & $\pm$154.9 & $\pm$143.6 & $\pm$33.5&     &  \\
16283$-$1613 & RST3950 & 26.23 & 2000.66 & 0.808 & 0.157 & 88.7 & 210.7 & 157.9 & 3 & Doc2009g \\
             &     & $\pm$0.17 & $\pm$0.35 & $\pm$0.018 & $\pm$0.006 & $\pm$16.1 & $\pm$16.6 & $\pm$12.8&     &  \\
16458$-$0046 & A 1141 & 31.15 & 2019.22 & 0.877 & 0.1140 & 302.1 & 106.9 & 164.0 & 2 & Baz1976 \\
             &     & $\pm$0.21 & $\pm$0.08 & $\pm$0.005 & $\pm$0.0034 & $\pm$21.5 & $\pm$20.5 & $\pm$8.8&     &  \\
16488$+$1039 & CRC 73 & 11.51 & 2021.16 & 0.176 & 0.200 & 27.6 & 187.5 & 71.4 & 3 & Tok2022f \\
             &     & $\pm$0.38 & $\pm$0.26 & $\pm$0.019 & $\pm$0.006 & $\pm$1.6 & $\pm$8.4 & $\pm$0.9&     &  \\
17415$-$5348 & HDS 2502 & 20.32 & 2018.045 & 0.593 & 0.1341 & 158.8 & 322.0 & 136.4 & 2 & Tok2019c \\
             &     & $\pm$0.22 & $\pm$0.027 & $\pm$0.006 & $\pm$0.0009 & $\pm$1.3 & $\pm$1.4 & $\pm$0.8&     &  \\
18480$-$1009 & HDS 2665 & 42.03 & 2023.906 & 0.313 & 0.4964 & 42.8 & 178.6 & 57.2 & 4 & Tok2019c \\  
             &     & $\pm$0.31 & $\pm$0.086 & $\pm$0.005 & $\pm$0.0019 & $\pm$0.5 & $\pm$1.8 & $\pm$0.4&     &  \\
18520$+$1358 & CHR 80 & 64.6 & 1990.18 & 0.662 & 0.161 & 98.5 & 280.2 & 121.4 & 3 & Tok2023d \\
             &     & $\pm$3.1 & $\pm$0.84 & $\pm$0.052 & $\pm$0.020 & $\pm$2.4 & $\pm$2.3 & $\pm$5.4&     &  \\
18520$-$5418 & TOK 325 Aa,Ab & 11.04 & 2017.45 & 0.241 & 0.1033 & 107.9 & 328.3 & 50.3 & 3 & Tok2022f \\
             &     & $\pm$0.19 & $\pm$0.11 & $\pm$0.015 & $\pm$0.0020 & $\pm$2.1 & $\pm$4.1 & $\pm$1.7&     &  \\
19035$-$6845 & FIN 357 & 14.102 & 2018.104 & 0.383 & 0.1493 & 148.8 & 234.7 & 155.5 & 2 & Doc2018i \\
             &     & $\pm$0.042 & $\pm$0.018 & $\pm$0.004 & $\pm$0.0010 & $\pm$3.5 & $\pm$3.8 & $\pm$1.4&     &  \\
19040$-$3804 & I 1391 & 48.75 & 2003.88 & 0.505 & 0.1868 & 123.4 & 194.7 & 52.3 & 3 & Tok2015c \\
             &     & $\pm$0.72 & $\pm$0.59 & $\pm$0.028 & $\pm$0.0022 & $\pm$3.0 & $\pm$6.1 & $\pm$2.6&     &  \\
19242$-$6300 & TOK 799 & 7.12 & 2024.08 & 0.585 & 0.088 & 127.3 & 356.2 & 122.6 & 3 & SOAR2023 \\
             &     & $\pm$0.46 & $\pm$0.12 & $\pm$0.075 & $\pm$0.005 & $\pm$3.1 & $\pm$7.9 & $\pm$4.3&     &  \\
19294$-$4057 & B 1385 & 51.82 & 1978.32 & 0.100 & 0.1821 & 126.5 & 108.7 & 50.5 & 2 & Tok2018b \\
             &     & $\pm$0.92 & $\pm$1.03 & $\pm$0.016 & $\pm$0.0015 & $\pm$1.5 & $\pm$4.1 & $\pm$0.9&     &  \\
19377$-$4128 & VOU 34 & 61.29 & 2001.24 & 0.065 & 0.1651 & 133.9 & 70.6 & 97.6 & 3 & Tok2017c \\
             &     & $\pm$1.95 & $\pm$2.78 & $\pm$0.043 & $\pm$0.0032 & $\pm$0.4 & $\pm$17.1 & $\pm$0.6&     &  \\
19453$-$6823 & TOK 425 Ba,Bb & 4.291 & 2017.128 & 0.85 & 0.0503 & 139.5 & 203.8 & 133.1 & 3 & Tok2019c \\
             &     & $\pm$0.072 & $\pm$0.079 & fixed & $\pm$0.0046 & $\pm$12.9 & $\pm$20.7 & $\pm$9.7&     &  \\
19563$-$3137 & TOK 698 & 12.79 & 2015.93 & 0.164 & 0.1085 & 59.6 & 44.0 & 90.8 & 3 & Tok2019c \\
             &     & $\pm$0.47 & $\pm$0.19 & $\pm$0.049 & $\pm$0.0040 & $\pm$0.4 & fixed & $\pm$0.5&     &  \\
20216$+$1930 & COU 327 AB & 51.7 & 1987.21 & 0.46 & 0.1306 & 67.5 & 326.1 & 81.0 & 3 & Doc2008a \\
             &     & $\pm$2.5 & $\pm$0.70 & $\pm$0.05 & $\pm$0.0037 & $\pm$0.9 & $\pm$11.6 & $\pm$1.5&     &  \\
20217$-$3637 & HDS 2908 & 12.699 & 2007.37 & 0.536 & 0.1113 & 107.9 & 336.2 & 81.8 & 2 & Tok2016e \\
             &     & $\pm$0.056 & $\pm$0.11 & $\pm$0.007 & $\pm$0.0016 & $\pm$0.4 & $\pm$3.2 & $\pm$0.6&     &  \\
20325$-$1637 & SEE 512 & 42.62 & 2001.79 & 0.915 & 0.136 & 139.0 & 199.6 & 130.9 & 2 & Doc2023 \\
             &     & $\pm$0.72 & $\pm$0.95 & $\pm$0.032 & $\pm$0.011 & $\pm$19.0 & $\pm$27.8 & $\pm$6.5&     &  \\
20527$-$0859 & MCA 64 Aa,Ab & 2.404 & 2018.973 & 0.150 & 0.0565 & 136.6 & 308.0 & 85.7 & 4 & Tok2023d \\  
             &     & $\pm$0.002 & $\pm$0.068 & $\pm$0.024 & $\pm$0.0008 & $\pm$0.7 & $\pm$9.8 & $\pm$1.2&     &  \\
21130$-$1133 & VOU 24 AB & 161.4 & 2013.66 & 0.316 & 0.309 & 98.0 & 250.0 & 143.2 & 4 & Tok2019c \\
             &     & $\pm$6.7 & $\pm$0.57 & $\pm$0.019 & $\pm$0.008 & $\pm$2.5 & fixed & $\pm$2.4&     &  \\
21278$-$5922 & TOK 731 & 12.99 & 2022.544 & 0.616 & 0.0600 & 198.0 & 128.6 & 27.6 & 3 & Tok2022g \\
             &     & $\pm$0.62 & $\pm$0.036 & $\pm$0.015 & $\pm$0.0013 & $\pm$10.0 & $\pm$11.4 & $\pm$4.2&     &  \\
21310$-$3633 & B 1008 AB & 75.05 & 2019.737 & 0.636 & 0.2222 & 32.3 & 163.3 & 102.9 & 4 & Tok2018b \\
             &     & $\pm$1.91 & $\pm$0.113 & $\pm$0.007 & $\pm$0.0015 & $\pm$0.3 & $\pm$1.5 & $\pm$0.4&     &  \\
21395$-$0003 & BU 1212 AB & 48.88 & 2020.814 & 0.861 & 0.4211 & 140.4 & 294.4 & 54.7 & 2 & Tok2019c \\
             &     & $\pm$0.10 & $\pm$0.007 & $\pm$0.001 & $\pm$0.0018 & $\pm$0.3 & $\pm$0.3 & $\pm$0.3&     &  \\
21477$-$3054 & FIN 330 AB & 20.23 & 2007.34 & 0.443 & 0.1232 & 32.5 & 216.5 & 108.8 & 2 & Tok2019c \\
             &     & $\pm$0.13 & $\pm$0.08 & $\pm$0.012 & $\pm$0.0013 & $\pm$0.7 & $\pm$1.8 & $\pm$0.4&     &  \\
21543$+$1943 & COU 432 BC & 62.9 & 2029.84 & 0.094 & 0.192 & 7.8 & 137.2 & 108.3 & 4 & Tok2022f \\
             &     & $\pm$2.1 & $\pm$2.79 & $\pm$0.037 & $\pm$0.008 & $\pm$1.3 & $\pm$17.7 & $\pm$1.3&     &  \\
22056$-$5858 & B 548 & 36.92 & 2025.27 & 0.624 & 0.148 & 35.9 & 337.0 & 56.3 & 2 & Tok2019c \\
             &     & $\pm$0.86 & $\pm$0.30 & $\pm$0.042 & $\pm$0.004 & $\pm$1.7 & $\pm$3.5 & $\pm$3.3&     &  \\
22441$+$0644 & TOK 703 & 4.887 & 2015.886 & 0.370 & 0.0588 & 170.5 & 343.3 & 135.4 & 3 & Tok2019c \\
             &     & $\pm$0.064 & $\pm$0.090 & $\pm$0.039 & $\pm$0.0026 & $\pm$6.8 & $\pm$12.8 & $\pm$5.4&     &  \\
22504$-$1744 & DON 1038 & 163.0 & 1992.09 & 0.432 & 0.3934 & 137.7 & 0.0 & 0.0 & 4 & Doc2016i \\
             &     & $\pm$5.6 & $\pm$0.61 & $\pm$0.017 & $\pm$0.0074 & $\pm$1.6 & fixed & fixed&     &  \\
23126$+$0241 & A 2298 AB & 29.161 & 2012.529 & 0.401 & 0.1989 & 287.0 & 312.9 & 102.7 & 2 & Pbx2000b \\
             &     & $\pm$0.054 & $\pm$0.092 & $\pm$0.005 & $\pm$0.0013 & $\pm$0.3 & $\pm$1.1 & $\pm$0.3&     &  \\
23179$-$0302 & YSC 167 & 15.89 & 2016.52 & 0.63 & 0.077 & 35.2 & 137.8 & 98.3 & 4 & Tok2023d \\
             &     & $\pm$0.61 & $\pm$1.80 & $\pm$0.16 & $\pm$0.028 & $\pm$4.7 & $\pm$41.4 & $\pm$4.6&     &  \\
23285$+$0926 & YSC 138 & 20.28 & 2008.12 & 0.539 & 0.0774 & 154.1 & 326.5 & 149.0 & 3 & Tok2022f \\
             &     & $\pm$0.73 & $\pm$0.17 & $\pm$0.036 & $\pm$0.0022 & $\pm$13.5 & $\pm$18.5 & $\pm$7.3&     &  \\
23350$+$0136 & MEL 9 BC & 35.23 & 2009.66 & 0.082 & 0.446 & 9.3 & 126.4 & 82.9 & 3 & MaN2019 \\
             &     & $\pm$0.94 & $\pm$1.18 & $\pm$0.012 & $\pm$0.012 & $\pm$0.4 & $\pm$13.2 & $\pm$0.2&     &  \\
\enddata 
\end{deluxetable*}



\startlongtable

\begin{deluxetable*}{l l cccc ccc cc}    
\tabletypesize{\scriptsize}     
\tablecaption{Preliminary Visual Orbits
\label{tab:vborb4}          }
\tablewidth{0pt}                                   
\tablehead{    
\colhead{WDS} & 
\colhead{Discoverer} & 
\colhead{$P$} & 
\colhead{$T$} & 
\colhead{$e$} & 
\colhead{$a$} & 
\colhead{$\Omega$ } & 
\colhead{$\omega$ } & 
\colhead{$i$ } & 
\colhead{Grade }  &
\colhead{Ref.} \\
\colhead{$\alpha$, $\delta$ (2000)} &
\colhead{Designation} & 
\colhead{(yr)} &
\colhead{(yr)} & &
\colhead{(arcsec)} & 
\colhead{(deg)} & 
\colhead{(deg)} & 
\colhead{(deg)} &  & 
}
\startdata
00036$-$3106 & TOK 686 & 14.014 & 2013.792 & 0.233 & 0.1310 & 11.8 & 264.9 & 71.2 & 4 & Tok2023d \\
00321$-$1218 & HDS 71 & 62.190 & 2035.840 & 0.216 & 0.2734 & 124.9 & 320.2 & 131.5 & 4 & Cve2017 \\
01503$-$8714 & HDS 247 & 100.000 & 2030.110 & 0.355 & 0.2056 & 90.8 & 92.5 & 115.4 & 4 & SOAR2023 \\
02035$-$0455 & TOK 873 & 20.000 & 2019.042 & 0.111 & 0.2303 & 51.8 & 196.8 & 28.3 & 4 & SOAR2023 \\
02143$-$4952 & TOK 815 & 9.196 & 2017.400 & 0.300 & 0.1446 & 273.4 & 111.7 & 78.0 & 4 & SOAR2023 \\
02297$-$0216 & BU 519 & 1000.000 & 2033.675 & 0.800 & 1.0282 & 80.5 & 104.5 & 64.9 & 3 & SOAR2023 \\
02370$-$3056 & RST 2283 & 143.078 & 1906.421 & 0.500 & 0.2762 & 33.8 & 54.0 & 53.8 & 4 & SOAR2023 \\
03417$-$5126 & HIP 17255 & 2.791 & 2021.688 & 0.170 & 0.0709 & 65.7 & 124.8 & 49.1 & 4 & Tok2023d \\
03571$-$0828 & RST 4764 & 200.000 & 1995.313 & 0.243 & 0.2270 & 155.9 & 353.9 & 124.7 & 4 & SOAR2023 \\
04237$+$1131 & LSC 30 & 72.551 & 2024.214 & 0.550 & 0.0861 & 52.2 & 58.8 & 50.0 & 4 & Tok2023d \\
04249$-$3445 & DAM 1313 Aa,Ab & 32.310 & 2021.200 & 0.500 & 0.1520 & 108.4 & 352.4 & 55.0 & 5 & SOAR2023 \\
04493$+$2934 & RAS 18 & 57.350 & 2026.540 & 0.174 & 0.1590 & 274.0 & 64.6 & 152.5 & 4 & SOAR2023 \\
04550$+$1436 & HDS 634 & 48.287 & 2013.029 & 0.527 & 0.1007 & 61.4 & 103.5 & 160.0 & 3 & SOAR2023 \\
05286$-$4548 & HDS 723 & 80.000 & 2016.917 & 0.937 & 0.1569 & 148.5 & 135.0 & 114.4 & 4 & SOAR2023 \\
05289$-$0318 & DA 6 & 1800.000 & 1995.856 & 0.860 & 1.0924 & 68.8 & 167.6 & 46.9 & 4 & Tok2015c \\
05321$-$0305 & JNN 39 Aa,Ab & 150.000 & 2019.445 & 0.859 & 0.7899 & 132.8 & 44.8 & 66.5 & 4 & Tok2022x \\
05321$-$0305 & JNN 39 Ba,Bb & 7.795 & 2022.159 & 0.000 & 0.0801 & 75.9 & 0.0 & 31.8 & 4 & SOAR2023 \\
05418$-$5000 & HU 1568 & 320.000 & 2019.186 & 0.327 & 1.0281 & 174.4 & 70.3 & 109.8 & 4 & Hrt2011d \\
06236$+$1739 & A 2517 & 300.000 & 1969.492 & 0.372 & 0.1866 & 26.7 & 139.0 & 105.5 & 4 & Tok2017c \\
07558$+$1320 & YSC 199 & 54.787 & 2008.175 & 0.000 & 0.1522 & 121.4 & 0.0 & 117.8 & 4 & SOAR2023 \\
08134$-$4534 & TOK 832 & 9.666 & 2018.994 & 0.000 & 0.0410 & 183.7 & 0.0 & 108.5 & 3 & SOAR2023 \\
09033$-$7036 & HEI 223 AB & 21.882 & 2023.220 & 0.206 & 0.0861 & 88.4 & 301.2 & 109.2 & 3 & Tok2022f \\
09095$-$5538 & YMG 29 & 12.000 & 2025.409 & 0.124 & 0.0478 & 157.2 & 212.9 & 39.4 & 3 & SOAR2023 \\
09133$-$5529 & ELP 24 & 33.172 & 2021.416 & 0.806 & 0.0949 & 124.8 & 163.3 & 150.5 & 4 & SOAR2023 \\
09180$-$5453 & JNN 69 Aa,Ab & 6.964 & 2020.846 & 0.381 & 0.0486 & 217.0 & 74.6 & 39.9 & 4 & Tok2023d \\
09477$+$2036 & COU 284 & 133.858 & 2068.313 & 0.880 & 0.4014 & 94.3 & 265.9 & 96.7 & 5 & Doc2019c \\
09494$+$1832 & HDS 1419 & 36.347 & 2021.487 & 0.962 & 0.1042 & 149.9 & 243.2 & 20.0 & 3 & SOAR2023 \\
09586$-$2420 & TOK 437 & 15.000 & 2012.429 & 0.223 & 0.0652 & 112.6 & 90.2 & 92.3 & 3 & SOAR2023 \\
10227$-$2350 & B 197 & 171.5 & 1986.73 & 0.684 & 0.216 & 134.3 & 212.6 & 20.0 & 4 & SOAR2023 \\
10356$-$4715 & GKM  Aa,Ab & 1.499 & 2016.643 & 0.471 & 0.0223 & 83.5 & 137.5 & 103.4 & 5 & SOAR2023 \\
10406$-$5342 & FIN 40 & 300.000 & 1927.108 & 0.200 & 0.3493 & 146.8 & 355.9 & 35.0 & 5 & SOAR2023 \\
10527$+$0029 & TOK 893 BC & 32.762 & 2001.454 & 0.199 & 0.2422 & 180.9 & 259.2 & 128.1 & 5 & SOAR2023 \\
11136$-$4749 & HDS 1602 & 100.000 & 2042.495 & 0.400 & 0.3276 & 37.2 & 15.6 & 84.2 & 5 & SOAR2023 \\
11430$-$3933 & RST 5358 & 151.850 & 2035.744 & 0.277 & 0.1466 & 168.2 & 125.0 & 160.0 & 4 & SOAR2023 \\
12175$+$0636 & BU 796 & 172.137 & 1980.853 & 0.290 & 0.3257 & 94.4 & 353.0 & 84.5 & 3 & SOAR2023 \\
12349$-$0509 & RST 4502 & 223.477 & 2006.343 & 0.200 & 0.2261 & 63.6 & 151.0 & 66.1 & 4 & SOAR2023 \\
12407$-$4803 & TOK 849 & 10.622 & 2018.485 & 0.500 & 0.0899 & 236.1 & 180.0 & 15.3 & 4 & SOAR2023 \\
13175$+$2024 & YSC 149 AB & 32.959 & 2016.098 & 0.085 & 0.2151 & 118.3 & 311.2 & 143.4 & 4 & SOAR2023 \\
13229$-$7209 & B 1736 & 73.039 & 1962.488 & 0.627 & 0.1628 & 156.4 & 324.4 & 54.2 & 4 & Tok2023d \\
13377$-$2337 & RST 2856 AB & 200.000 & 2117.606 & 0.046 & 0.3502 & 103.6 & 15.6 & 79.9 & 3 & Tok2016e \\
13401$-$6033 & TOK 292 Ca,Cb & 40.000 & 2019.059 & 0.267 & 0.1987 & 131.9 & 195.4 & 155.5 & 4 & Tok2023d \\
14490$-$5807 & HDS 2087 & 28.815 & 2027.684 & 0.320 & 0.1670 & 111.7 & 14.4 & 121.4 & 3 & Tok2023d \\
16076$+$0002 & HDS 2276 & 161.761 & 2024.282 & 0.774 & 0.6006 & 172.0 & 172.1 & 71.5 & 5 & Tok2018i \\ 
16120$-$1928 & CHR 146 Aa,Ab & 10.061 & 2014.987 & 0.920 & 0.0733 & 346.4 & 118.9 & 89.7 & 3 & Tok2021f \\ 
16161$-$3037 & I 1586 AB & 180.000 & 2110.616 & 0.129 & 0.3347 & 145.6 & 133.4 & 154.6 & 4 & Tok2019c \\
16385$+$1240 & TOK 727 & 8.647 & 2022.391 & 0.492 & 0.1009 & 171.8 & 316.0 & 92.9 & 3 & Tok2023d \\
16536$-$1045 & YSC 156 & 31.287 & 2024.604 & 0.299 & 0.1076 & 25.4 & 129.7 & 35.6 & 3 & Tok2020g \\
17184$-$0147 & BAG 51 & 30.372 & 2000.506 & 0.200 & 0.4637 & 121.5 & 309.8 & 58.2 & 4 & Tok2023d \\
17535$-$0355 & TOK 54 & 47.309 & 2032.612 & 0.242 & 0.0928 & 151.1 & 0.0 & 180.0 & 4 & Tok2023d \\
19301$-$4904 & HDS 2772 & 290.000 & 2019.614 & 0.823 & 0.5913 & 175.9 & 111.2 & 64.4 & 4 & Tok2017c \\
19369$-$6949 & GKM  Aa,Ab & 2.013 & 2015.363 & 0.154 & 0.0239 & 208.1 & 42.8 & 142.3 & 4 & SOAR2023 \\
19561$-$3208 & BWL 53 Ba,Bb & 20.000 & 2025.372 & 0.522 & 0.1529 & 33.6 & 45.6 & 55.0 & 3 & Tok2023a \\
20205$-$2749 & RST 3255 & 54.643 & 2027.239 & 0.929 & 0.1011 & 266.9 & 0.0 & 0.0 & 3 & Hei1996a \\ 
22006$-$1345 & HU 282 & 500.000 & 2029.823 & 0.900 & 0.4846 & 102.4 & 93.1 & 25.1 & 4 & Tok2023d \\
22343$+$0345 & HDS 3201 & 27.770 & 2000.161 & 0.451 & 0.1221 & 36.8 & 98.0 & 52.0 & 3 & Tok2023d \\ 
23025$-$4605 & I 1462 & 155.000 & 2035.714 & 0.619 & 0.1943 & 105.0 & 282.4 & 160.0 & 4 & Tok2023d \\
23052$-$1822 & B 1898 & 119.895 & 1912.955 & 0.406 & 0.2569 & 236.3 & 94.2 & 116.8 & 4 & SOAR2023 \\
23210$-$0229 & LSC 107 & 35.000 & 2015.328 & 0.575 & 0.1327 & 176.1 & 13.4 & 27.4 & 4 & Tok2022f \\
23224$-$6516 & HDS 3328 & 45.095 & 2022.798 & 0.679 & 0.1432 & 110.2 & 134.9 & 69.3 & 3 & Tok2022f \\
23384$-$2922 & B 606 & 182.729 & 2035.648 & 0.000 & 0.3177 & 175.7 & 0.0 & 83.7 & 4 & SOAR2023 \\
\enddata 
\end{deluxetable*}


\subsection{Combined Spectro-Interferometric Orbits}
\label{sec:combined}

\begin{figure*}[ht]
\epsscale{1.1}
\plotone{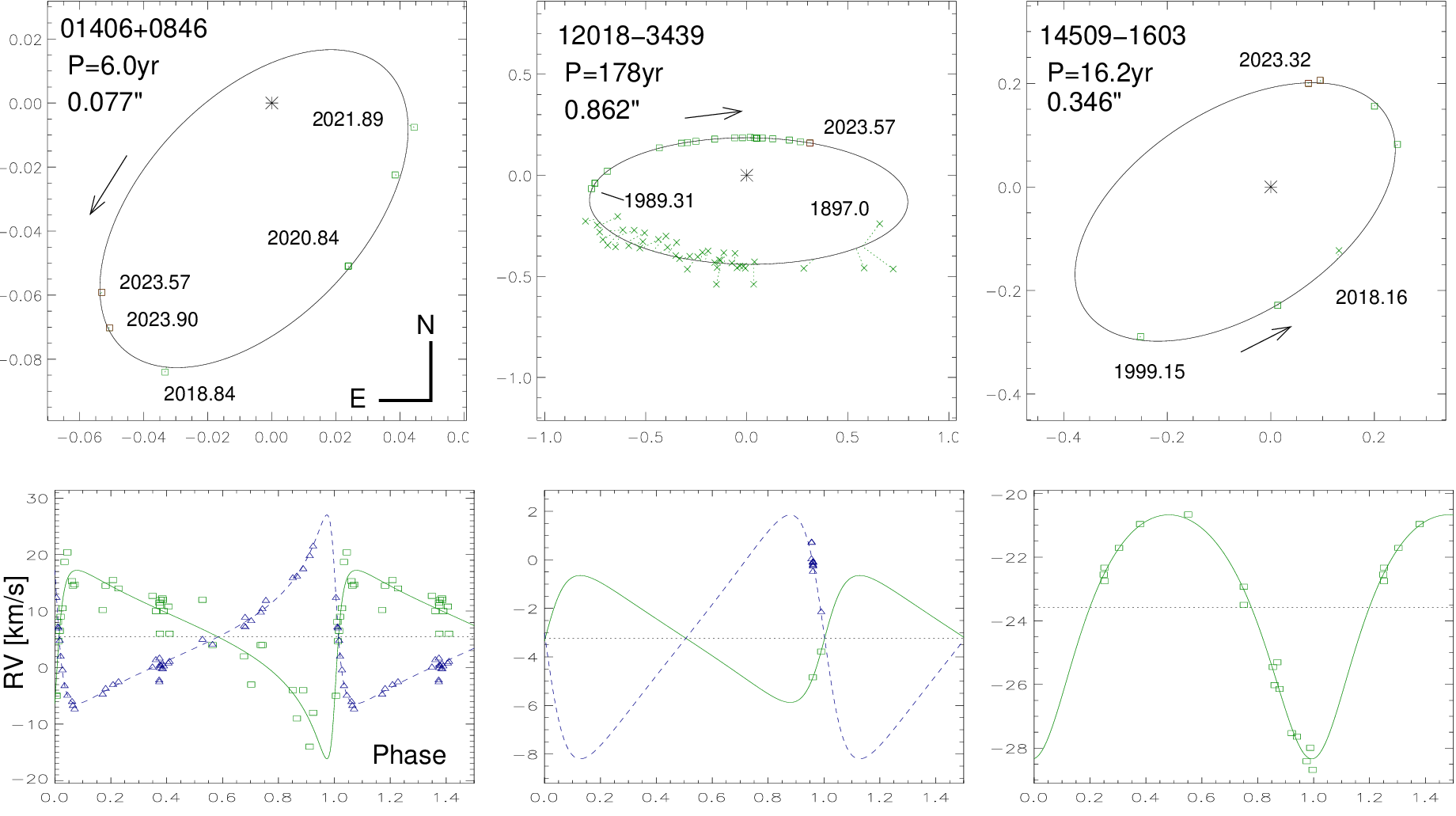}
\caption{Three combined  spectro-interferometric orbits. The  upper plots
  are similar  to those in Figure~\ref{fig:visual},  the corresponding
  RV curves are shown below.
\label{fig:combined} }
\end{figure*}

In calculation  of eight orbits  we used the radial  velocities (RVs),
fitting  combined spectro-visual  orbits.  In  addition to  the visual
elements in Table~\ref{tab:vborb}, we provide in Table~\ref{tab:sborb}
the RV amplitudes of the primary and secondary components $K_1$, $K_2$
and the systemic  velocity $V_0$.  These parameters,  delivered by the
combined fits  of positions  and RVs,  are similar but  not equal  to the
published spectroscopic elements.  The  first three columns repeat the
WDS codes, DDs,  and periods, and the last column  gives references to
publications containing the RVs used here. In the combined orbits, the
angles $\Omega$ and $\omega$ are  selected to match the ascending node
and RV  of the  primary component  and the  relative positions  of the
secondary component.

\begin{deluxetable*}{l l cc cc l}    
\tabletypesize{\scriptsize}     
\tablecaption{Combined Spectro-Interferometric Orbits
\label{tab:sborb}          }
\tablewidth{0pt}                                   
\tablehead{                                                                     
\colhead{WDS} & 
\colhead{Discoverer} & 
\colhead{$P$} & 
\colhead{$K_1$} & 
\colhead{$K_2$} & 
\colhead{$V_0$} &
\colhead{Reference}  \\
\colhead{$\alpha$, $\delta$ (2000)} &
\colhead{Designation} & 
\colhead{(yr)} &
\colhead{(\kms)} &
\colhead{(\kms)} &
\colhead{(\kms)} & 
}
\startdata
01406$+$0846 & TOK 872 & 5.979 & 16.69 & 16.70 & 5.43 &  \citet{Griffin2007} \\ 
             &   & $\pm$0.026 & $\pm$1.08 & $\pm$0.28 & $\pm$0.11 &  \\
12018$-$3439 & I 215 AB & 178.48 & 2.61 & 5.02 & $-$3.23 &  \citet{Tok2015} \\ 
             &   & $\pm$4.50 & $\pm$0.31 & $\pm$0.31 & $\pm$0.14 &  \\
12114$-$1647 & S 634 Aa,Ab & 0.5792 & 15.48 & 17.71 & 2.35 &  \citet{CHIRON6} \\
             &   & $\pm$0.0006 & $\pm$0.04 & $\pm$0.04 & $\pm$0.02 &  \\
13133$+$1621 & DOC 1 & 25.44 & 6.04 & 6.53 & $-$7.01 &  \citet{Griffin2017} \\ 
             &   & $\pm$0.17 & $\pm$0.13 & $\pm$0.13 & $\pm$0.09 &  \\
13513$-$2423 & WSI 77 & 10.490 & 6.18 & \ldots & 5.27 &  \citet{Willmarth2016} \\ 
             &   & $\pm$0.006 & $\pm$0.04 & \ldots & $\pm$0.03 &  \\
14509$-$1603 & BEU 19 Ba,Bb & 16.229 & 3.83 & \ldots & $-$23.57 &  \citet{DM91} \\ 
             &   & $\pm$0.043 & $\pm$0.12 & \ldots & $\pm$0.09 &  \\
15006$+$0836 & YSC 8 & 6.922 & 10.19 & 10.23 & 8.13 &  \citet{Halbwachs2020} \\ 
             &   & $\pm$0.003 & $\pm$0.05 & $\pm$0.04 & $\pm$0.02 &  \\
23126$+$0241 & A 2298 AB & 29.161 & 5.24 & 8.04 & 34.48 &  \citet{Pourbaix2000} \\ 
             &   & $\pm$0.054 & $\pm$0.14 & $\pm$0.25 & $\pm$0.11 &  \\
\enddata 
\end{deluxetable*}

Three       combined      spectro-interferometric       orbits      in
Figure~\ref{fig:combined} illustrate different contributions of the RV
data to  the combined solutions. The  double-lined spectroscopic orbit
of 01406+0846 (HD  10262, TOK 872, spectral type F2)  with a period of
5.85 yr was published by  \citet{Griffin2007}.  The RVs from his paper
strongly constrain the orbital period.   However, the RV amplitudes of
16.0 and 16.1 \kms derived  by Griffin, together with the inclination,
correspond to  the nearly  equal component's masses  of 2.24  \msun and
contradict the  substantial magnitude difference  of $\Delta y  = 1.2$
mag measured  at SOAR.  The large  scatter in the RV  curve attests to
the uncertain splitting  of the blended lines used to  derive the RVs.
Spectroscopy with a higher resolution  and future less biased parallax
from Gaia  that would explicitly  account for the orbital  motion will
eventually lead to  accurate measurement of the  masses.  All position
measures  come  from  SOAR;  it   was  unresolved  in  2022  near  the
periastron, when the secondary moved to the opposite quadrant.

The second example in Figure~\ref{fig:combined}, 12018$-$3439 (I 215, HD
104471) is a classical  long-period visual pair. Spectroscopy revealed
this  as a  triple-lined  star containing  a  148-day inner  subsystem
\citep{Tok2015}. The RVs of both visual components (center of mass of
the  inner  binary   and  the  visual  secondary)   help  somewhat  in
constraining the eccentricity.  Here, the century-long historic visual
data, although less accurate than  speckle, are essential in providing
nearly complete  coverage of the  orbit.  Further improvement  of this
orbit will be  painfully slow.  The third example,  14509$-$1603 (BEU 19
Ba,Bb,  HD  130819,  $\alpha^1$~Lib),   refers  to  the  spectroscopic
subsystem  paired to  the bright  star $\alpha^2$~Lib  (HD 130841)  at
231\arcsec, which is itself a 70-day spectroscopic binary. The RVs of Ba
are taken  from the seminal  paper by \citet{DM91}.  The  pair Ba,Bb
has been resolved  by J.-L.  Beuzit  in 2004,  while all other  measures come
from  SOAR.  The  magnitude difference  is  $\sim$4.5 mag  in the  $I$
filter, so the measures are less accurate than usual.

Combined   visual-spectroscopic  orbits   allow  measurement   of  the
component's masses  and distances  independently of  the trigonometric
parallax  \citep[e.g.][]{Pourbaix2000}.    However,  the   masses  are
proportional  to the  cube of  the RV  amplitudes and  to $\sin^3  i$,
therefore high  accuracy of these  parameters is required  for getting
usefully accurate masses.  The example  of 01406+0846 above shows that
the small RV amplitudes cannot be trusted, and the masses based on the
combined orbit  are misleading.  On  the other hand,  the double-lined
spectroscopic orbit of 12114$-$1647 with a period of 211 days is quite
accurate, but  its semimajor axis  of 25.3\,mas,  at the limit  of the
SOAR resolving power, is too  small for measuring accurate masses (the
inclination   is    poorly   constrained).     The   only    pair   in
Table~\ref{tab:sborb} with  sufficiently accurate speckle and  RV data
is  15006+0836  (HIP  73449);   the  masses  are  0.902$\pm$0.008  and
0.899$\pm$0.008     \msun    and     the    orbital     parallax    is
26.31$\pm$0.15\,mas. The  GDR3 parallax  of 26.26$\pm$0.16\,mas  is in
agreement because the components are  nearly equal and the photocenter
is little affected by the orbital motion. This is not the case for the
majority of close visual pairs with absent or biased GDR3 parallaxes.




\subsection{Hierarchies within 100 pc}
\label{sec:GKM}

\begin{figure*}[ht]
\epsscale{0.7}
\plotone{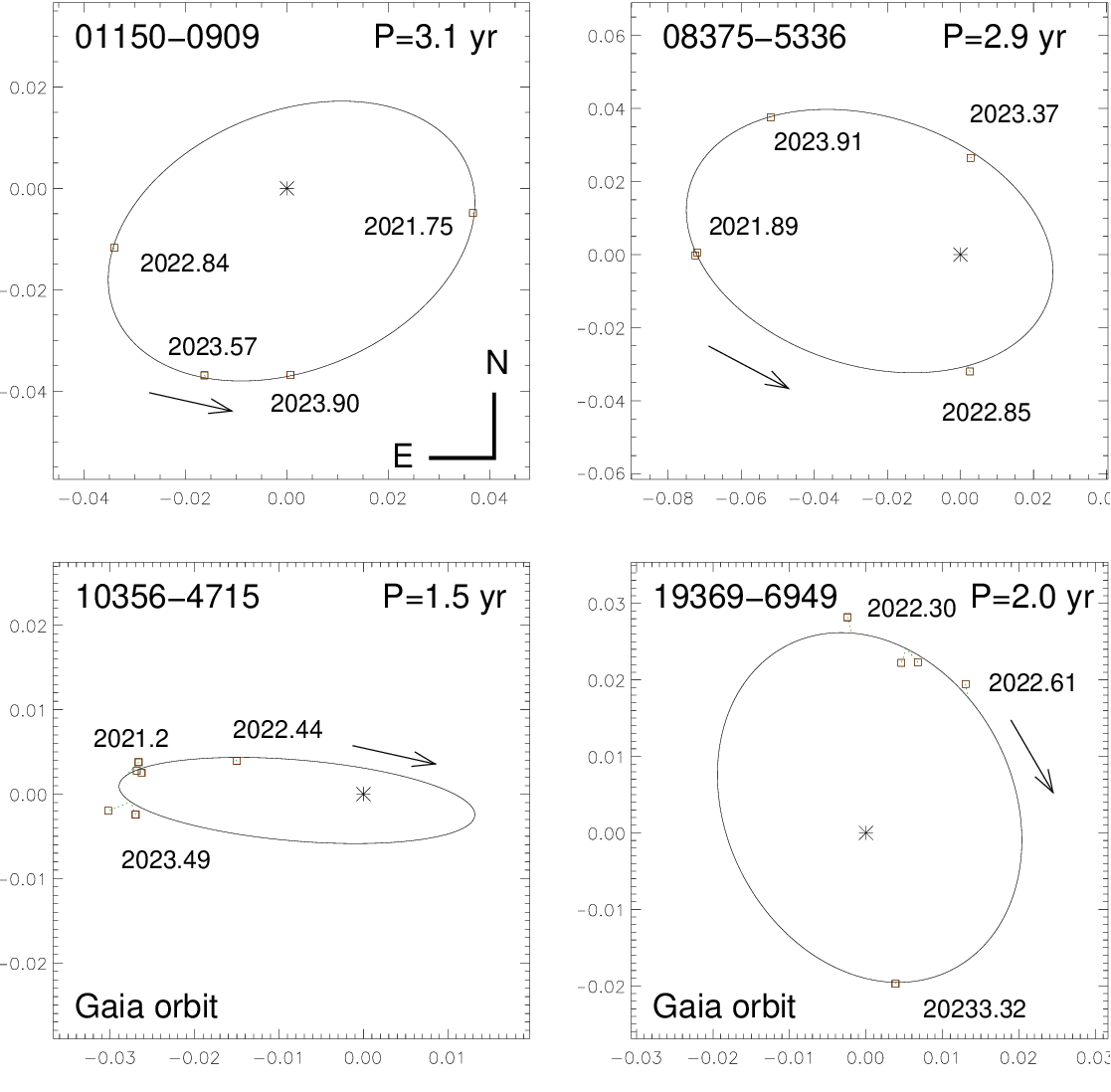}
\caption{Orbits of four inner pairs in nearby hierarchical
  systems. The axis scale is in arcseconds.
\label{fig:GKMorbits} }
\end{figure*}

\begin{figure}[ht]
\epsscale{1.1}
\plotone{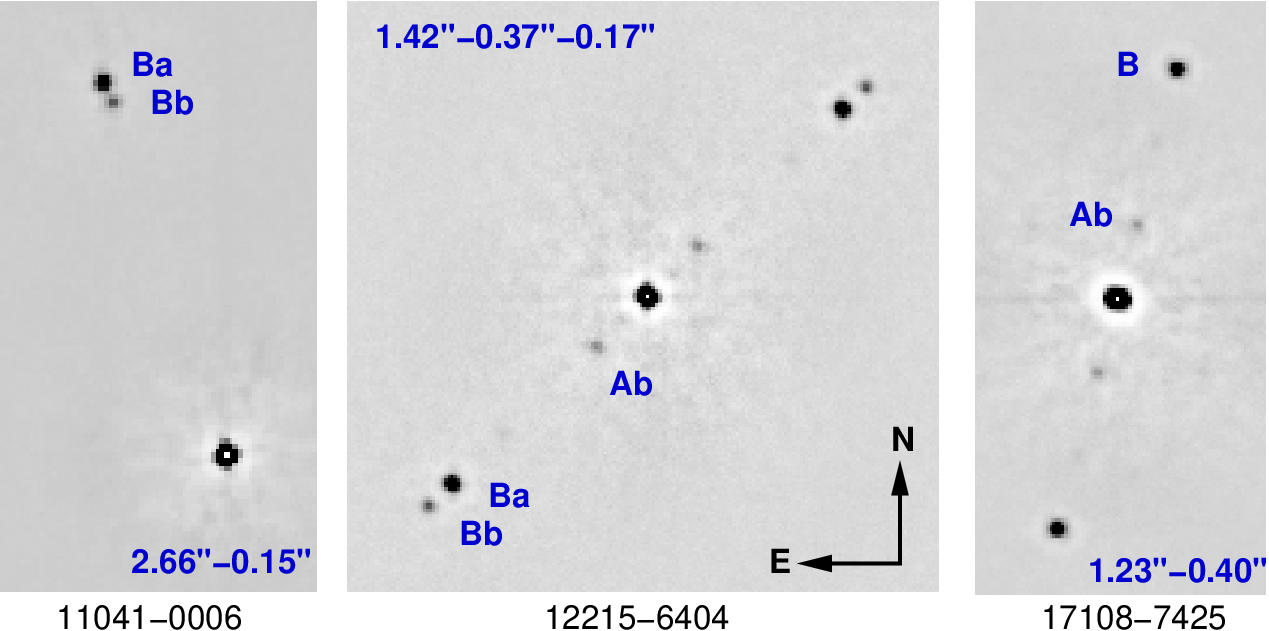}
\caption{Fragments  of speckle  auto-correlation  functions (ACFs)  of
  three  newly  identified  compact  hierarchies  within  100  pc  (in
  arbitrary  negative  rendering).  Outer and  inner  separations  are
  indicated,  and  the  ACF  peaks  corresponding  to  true  companion
  locations are labeled.
\label{fig:GKMtrip} }
\end{figure}

During 2021-2022, HRCam has been used to probe components of 1200 wide
binaries  located within  100\,pc that  had indications  of unresolved
inner subsystems  in the Gaia data;  about half of those  targets were
actually resolved, as reported by \citet{GKM}.  The remaining half are
likely beyond  the speckle  detection limit,  either too  close and/or
with large  contrast. In 2023 we revisited some  of those new pairs,
mostly the closest ones. The goal was twofold. First, to determine the
sense of motion which, when combined with the Gaia astrometry of the
outer  components,  gives statistical  information  on  the relative  orbit
orientation.   Second,  to collect  data  for  determination of  inner
orbits.
 
Figure~\ref{fig:GKMorbits}  presents the  first preliminary  orbits of
inner  subsystems  in  nearby   hierarchies  derived  from  the  HRCam
data. Two  pairs in the top  row have four measurements  each covering
most of their  3-yr orbits.  The two remaining  pairs have astrometric
and spectroscopic  orbits in the GDR3,  and we fixed some  elements to 
those  values while  fitting  the speckle  measurements.  Elements  of
these   orbits   are  given   in   the   Tables  \ref{tab:vborb}   and
\ref{tab:vborb4}.  Continued speckle  monitoring of nearby hierarchies
from this program will yield more inner orbits in the near future.

The original  100-pc program,  started in  2021, targeted  pairs wider
than 3\arcsec ~with  reliable Gaia astrometry of  both components.  It
was  later  extended  to  pairs with  separations  from  1\arcsec  ~to
3\arcsec,  and  more such  pairs  were  observed  during 2023  at  low
priority.  In ten  pairs, the expected inner  subsystems were actually
resolved,  and  one   of  them  was  revealed  as   a  2+2  quadruple.
Representative  results are  shown  in Figure~\ref{fig:GKMtrip}.   The
first triple, 11041$-$0006, is  quite typical: its secondary component
was resolved into  a 0\farcs15 pair Ba,Bb; this system  with the outer
separation of  2\farcs66 is  fairly hierarchical.  Most  newly resolved
triples  are  similarly  hierarchical.   However,  the  2+2  quadruple
12215$-$6404 (FP Gru, K3.5e) illustrated in the middle panel is not so
hierarchical, the  ratio of separations  in the outer Aa,Ba  and inner
Aa,Ab  pairs  is  only  3.8,  not far  from  the  limit  of  dynamical
stability.  Interestingly, this quadruple  has a linear configuration,
suggesting coplanar orbits  if it is viewed edge-on.   These stars are
chromospherically  active   and  likely  young.   The   triple  system
17108$-$7425 (HIP  84035, G2V)  in the right-hand  panel is  even more
extreme  in terms  of its  separation  ratio of  3.1, and  it is  also
arranged linearly;  the projected separations suggest  periods between
100 and  600 yr.  There  is also a  common proper motion  companion at
734\arcsec  (LDS 582  AC) with a projected separation  of 53\,kau  that
might be either gravitationally bound or simply co-moving.

Weakly   hierarchical  systems   of  low-mass   stars  in   the  solar
neighborhood  are  amenable to  the  study  of their  dynamics,  while
explaining  their origin  challenges  star  formation theory.  The
prototype of such systems, LHS~1070 (00247$-$2653), is being monitored
at SOAR, and  deviations from the Keplerian motion in  the inner orbit
caused by interaction with the outer star are measurable. The SOAR speckle
program has revealed several more  weak low-mass hierarchies, to which
we add  two here.

\subsection{Spurious Pairs}
\label{sec:bogus}

\startlongtable

\begin{deluxetable}{ l l l l  } 
\tabletypesize{\scriptsize}    
\tablecaption{Likely Spurious Pairs
\label{tab:bogus} }                   
\tablewidth{0pt}     
\tablehead{ \colhead{WDS}  &
\colhead{Discoverer}  &  
\colhead{Resolved} & 
\colhead{Unresolved\tablenotemark{a}} 
}
\startdata  
00374$-$3717 & I 705       & 0\farcs4  Vis 1910 & 2008--23, S, R  \\ 
00558$-$1832 & B 645         & 0\farcs2  Vis 1926 & 2008--23,  R  \\ 
05484+0745  & JNN 267      & 1\farcs6  Spe 2011 & 2023, L  \\ 
09407$-$6639 & TDS6731       & 0\farcs5  Tycho    & 2023, R  \\ %
10328+0918 & WRH 19        & 0\farcs1  Vis 1937 & 2009--23  \\ 
10341+1222 & CHR 31        & 0\farcs2  Spe 1983 & 2023 \\ 
11006+0337 & CHR 33        & 0\farcs2  Spe 1983 & 2014--23, S, R    \\ 
11479+0815 & CHR 134 Aa,Ab & 0\farcs3  Spe 1987 & 2014--23, S, R \\ 
11518$-$0546 & CHR 36        & 0\farcs2  Spe 1983 & 2014--23, L, R  \\ 
15355$-$1447 & WRH 20 Aa,Ab  & 0\farcs1  Vis 1937 & 2009--23  \\ 
16133+1332 & CHR 52 Aa,Ab  & 0\farcs2  Spe 1983 & 2008--23, S, R   \\ 
17376$-$1524 & ISO 6 Aa,Ab   & 0\farcs3  Spe 1987 & 2023  \\ 
18073+0934 & STT 342 Aa,Ab & 1\farcs5  Vis 1842 & 2023 \\ 
19247+0833 & WSI 108       & 0\farcs1  Spe 2008 & 2015--23, S, R  \\ 
19255+0307 & BNU 6 Aa,Ab   & 0\farcs1  Spe 1979 & 2015--23 \\ 
20285$-$2410 & CHR 98        & 0\farcs2  Spe 1983 & 2014--23 \\ 
21400+0911 & CHR 105 AB    & 0\farcs3  Spe 1982 & 2008--23  \\ 
23157+0118 & CHR 141 AB    & 0\farcs1  Spe 1986 & 2013-23, S \\ 
\enddata
\tablenotetext{a}{Additional indications of the spurious nature of visual pairs:
R -- no excess noise in GDR3, RUWE$<$2; 
L -- long estimated period; 
S -- short estimated period or spectroscopic coverage.
}
\end{deluxetable}

Some  double stars  listed  in  the WDS  are  actually single.   These
spurious pairs  originate from  erroneous visual  resolutions, dubious
speckle data caused by instrumental  artifacts, and for other reasons.
Identifying spurious pairs  will save observing time in  the future by
eliminating  the  need to  followup  and  examine these  targets.   In
Table~\ref{tab:bogus} are listed pairs we identify as likely spurious,
continuing the clean-up effort from our previous papers. In this table
we provide the  WDS identifier and DD, the method
and date of the original discovery, and  the year(s) it has been unresolved
in this  program.  Following that  is a code giving  other indications
supporting the  characterization of the double  as spurious \citep[see
  details in][]{SAM21}.  In  the WDS \citep{WDS}, these  pairs are not
removed but are given an {\bf  X} code identifying them as a ``dubious
double'' or a ``bogus binary''.


\section{Summary and outlook}
\label{sec:sum}

Binary  stars observed  in this  program illustrate  progress in  this
field, from historic visual discoveries to Hipparcos and speckle pairs
resolved at the  end of the 20th century and  to the modern, sometimes
quite recent, resolutions.  The Gaia mission becomes a major driver of
speckle programs,  in a similar role  as played by Hipparcos,  but on a
much greater scale.

The  GDR3  data   already  include  a   million    new  pairs
\citep{ElBadry2021},  almost  an  order  of magnitude  more  than  the
content of the  WDS. However, these wide Gaia pairs  move slowly
and are  not amenable  to calculation  of orbits  in the near future.  The
historic micrometer pairs, in contrast,  have a larger time coverage
and, despite  the low  accuracy of old  measures, their  monitoring at
SOAR  steadily  produces  new   orbital  solutions.   Eventually,  the
resource  of historic  pairs will  be exhausted,  although incremental
improvement of their orbits will continue  for a long time, until they
are fully covered by accurate measures. 

An increasing  number of new visual  orbits is being computed  for the
tight and fast pairs discovered  recently by speckle.  The $10^5$ Gaia
astrometric  orbits outnumber  the  catalog of  visual  orbits by  two
orders of magnitude.  Future Gaia  data releases will further increase
the number of astrometric orbits and will extend their maximum periods
from 3 to 10 years. However, actual resolution of astrometric binaries
by speckle  interferometry is needed  for measurement of  their masses
(see  Figure~\ref{fig:dmsep}).  Two  relevant  examples  are shown  in
lower panel  of Figure~\ref{fig:GKMorbits}. Furthermore, the  range of
periods  from 10  to 100  yr will  not be  accessible to  Gaia without
complementary ground-based data, and obtaining this data sooner rather
than later will provide the required time coverage.  This period range
corresponds  to the  maximum of  the period  distribution of  low-mass
stars and matches the periods of  massive planets in the solar system.
The match is  not coincidental because formation of  both binaries and
planets is related to the  typical size of circumstellar disks, 10-100
au.

The Gaia catalog  of nearby stars within 100\,pc  contains 61644 stars
brighter than $G=12$  mag; 16315 of those (26\%)  are likely binaries,
based on the increased astrometric  noise or double transits.  If half
of  these candidates  can be  resolved by  speckle interferometry,  as
indicated by the recent survey \citep{GKM}, there are potentially 8000
targets  for future  orbits. For  comparison, the  WDS contains  about
12600 pairs  closer than  0\farcs5, but  most of  these pairs  are too
distant and  too slow for  orbit determination.  So, Gaia  becomes the
major driver of future speckle programs and future work on orbits

Considering the large number of potential speckle targets, one may ask
whether it is necessary to observe  all of them. A much smaller sample
of {\em  accurate} orbits and  parallaxes will suffice to  address the
classical  issue of  stellar  masses and  the  calibration of  stellar
evolutionary models. Current  projects in this area focus  on the less
explored  mass ranges,  i.e.   small \citep{Mann2019,Vrijmoet2022}  or
large masses. However,  accurate orbits of massive,  close, and bright
pairs are  better determined by long-baseline  interferometers than by
speckle.  But the  role of visual orbits is much  wider than just mass
measurements.  The  SOAR speckle program concentrates  on the dynamics
of  stellar  hierarchical  systems \citep[e.g.][]{TRI23}.   Orbits  of
exoplanet hosts \citep{Lester2023} or of  very young stars are other
interesting research  areas. Looking  into the  future, we  foresee an
increased  demand for  speckle astrometry  of binary  stars driven  by
diverse astrophysical applications and stimulated by Gaia.

\begin{acknowledgments} 

We thank the SOAR operators for efficient support of this program, and
the SOAR director J.\ Elias for allocating some technical time.  R.A.M
and E.C acknowledge support from  the Vicerrectoria de Investigacion y
Desarrollo (VID) de la Universidad de Chile, project number ENL02/23,
and from the FONDECYT grant number  1240049. The research of A.T.\ is
supported by the NSFs NOIRLab.

This work used the SIMBAD service operated by Centre des Donn\'ees Stellaires 
(Strasbourg, France), bibliographic references from the Astrophysics Data System
maintained by SAO/NASA, and the Washington Double Star Catalog maintained at the
USNO. This work has made use of data from the European Space Agency (ESA) 
mission Gaia (\url{https://www.cosmos.esa.int/gaia}) processed by the Gaia Data
Processing and Analysis Consortium (DPAC, {\url 
https://www.cosmos.esa.int/web/gaia/dpac/consortium}). Funding for the DPAC has 
been provided by national institutions, in particular the institutions 
participating in the Gaia Multilateral Agreement.

\end{acknowledgments} 


\facility{SOAR}


\bibliography{soar.bib}
\bibliographystyle{aasjournal}



\end{document}